\documentclass[twoside,twocolumn,9pt]{article}
\usepackage{extsizes}
\usepackage[super,sort&compress,comma]{natbib} 
\usepackage[version=3]{mhchem}
\usepackage[left=1.5cm, right=1.5cm, top=1.785cm, bottom=2.0cm]{geometry}
\usepackage{balance}
\usepackage{times,mathptmx}
\usepackage{sectsty}
\usepackage{graphicx} 
\usepackage{lastpage}
\usepackage[format=plain,justification=justified,singlelinecheck=false,font={stretch=1.125,small,sf},labelfont=bf,labelsep=space]{caption}
\usepackage{float}
\usepackage{fancyhdr}
\usepackage{fnpos}
\usepackage[english]{babel}
\usepackage{array}
\usepackage{droidsans}
\usepackage{charter}
\usepackage[T1]{fontenc}
\usepackage[usenames,dvipsnames]{xcolor}
\usepackage{setspace}
\usepackage[compact]{titlesec}
\usepackage[pdftitle={A ferroelectric liquid crystal confined in cylindrical nanopores: Reversible smectic layer buckling, enhanced light rotation and extremely fast electro-optically active Goldstone excitations}, pdfauthor={Mark Busch, Andriy V. Kityk, Wiktor Piecek, Tommy Hofmann, Dirk Wallacher, Sylwia Calus, Przemyslaw Kula, Martin Steinhart, Manfred Eich and Patrick Huber}]{hyperref}

\usepackage[euler]{textgreek}

\usepackage{epstopdf}

\definecolor{cream}{RGB}{222,217,201}

\begin{document}

\pagestyle{fancy}
\thispagestyle{plain}
\fancypagestyle{plain}{

\renewcommand{\headrulewidth}{0pt}
}

\makeFNbottom
\makeatletter
\renewcommand\LARGE{\@setfontsize\LARGE{15pt}{17}}
\renewcommand\Large{\@setfontsize\Large{12pt}{14}}
\renewcommand\large{\@setfontsize\large{10pt}{12}}
\renewcommand\footnotesize{\@setfontsize\footnotesize{7pt}{10}}
\makeatother

\renewcommand{\thefootnote}{\fnsymbol{footnote}}
\renewcommand\footnoterule{\vspace*{1pt}%
\color{cream}\hrule width 3.5in height 0.4pt \color{black}\vspace*{5pt}} 
\setcounter{secnumdepth}{5}

\makeatletter 
\renewcommand\@biblabel[1]{#1}            
\renewcommand\@makefntext[1]%
{\noindent\makebox[0pt][r]{\@thefnmark\,}#1}
\makeatother 
\renewcommand{\figurename}{\small{Fig.}~}
\sectionfont{\sffamily\Large}
\subsectionfont{\normalsize}
\subsubsectionfont{\bf}
\setstretch{1.125} 
\setlength{\skip\footins}{0.8cm}
\setlength{\footnotesep}{0.25cm}
\setlength{\jot}{10pt}
\titlespacing*{\section}{0pt}{4pt}{4pt}
\titlespacing*{\subsection}{0pt}{15pt}{1pt}

\fancyfoot{}
\fancyfoot[RO]{\footnotesize{\sffamily{\thepage}}}
\fancyfoot[LE]{\footnotesize{\sffamily{\thepage}}}
\fancyhead{}
\renewcommand{\headrulewidth}{0pt} 
\renewcommand{\footrulewidth}{0pt}
\setlength{\arrayrulewidth}{1pt}
\setlength{\columnsep}{6.5mm}
\setlength\bibsep{1pt}

\makeatletter 
\newlength{\figrulesep} 
\setlength{\figrulesep}{0.5\textfloatsep} 

\newcommand{\topfigrule}{\vspace*{-1pt}%
\noindent{\color{cream}\rule[-\figrulesep]{\columnwidth}{1.5pt}} }

\newcommand{\botfigrule}{\vspace*{-2pt}%
\noindent{\color{cream}\rule[\figrulesep]{\columnwidth}{1.5pt}} }

\newcommand{\dblfigrule}{\vspace*{-1pt}%
\noindent{\color{cream}\rule[-\figrulesep]{\textwidth}{1.5pt}} }

\makeatother

\twocolumn[
  \begin{@twocolumnfalse}
\sffamily

\noindent\LARGE{\textbf{A ferroelectric liquid crystal confined in cylindrical nanopores: Reversible smectic layer buckling, enhanced light rotation and extremely fast electro-optically active Goldstone excitations}} \\

\noindent\large{Mark Busch,\textit{$^{a}$} Andriy V.\ Kityk,$^{\ast}$\textit{$^{a,b}$} Wiktor Piecek,\textit{$^{c}$} Tommy Hofmann,\textit{$^{d}$} Dirk Wallacher,\textit{$^{d}$} Sylwia Ca{\l}us,\textit{$^{b}$} Przemys{\l}aw Kula,\textit{$^{c}$} Martin Steinhart,\textit{$^{e}$} Manfred Eich\textit{$^{f,g}$} and Patrick Huber$^{\ast}$\textit{$^{a}$}} \\
 
\begin{tabular}{m{1.5cm} p{14.5cm} m{1.5cm}}

 & \noindent\normalsize{The orientational and translational order of a thermotropic ferroelectric liquid crystal (2MBOCBC) imbibed in self-organized, parallel, cylindrical pores with radii of 10, 15, or 20~nm in anodic aluminium oxide monoliths (AAO) are explored by high-resolution linear and circular optical birefringence as well as neutron diffraction texture analysis. The results are compared to experiments on the bulk system. The native oxidic pore walls do not provide a stable smectogen wall anchoring. By contrast, a polymeric wall grafting enforcing planar molecular anchoring results in a thermal-history independent formation of smectic~C* helices and a reversible chevron-like layer buckling. An enhancement of the optical rotatory power by up to one order of magnitude of the confined compared to the bulk liquid crystal is traced to the pretransitional formation of helical structures at the smectic-A*-to-smectic-C* transformation. A linear electro-optical birefringence effect evidences collective fluctuations in the molecular tilt vector direction along the confined helical superstructures, \textit{i.e.}\ the Goldstone phason excitations typical of the para-to-ferroelectric transition. Their relaxation frequencies increase with the square of the inverse pore radii as characteristic of plane-wave excitations and are two orders of magnitude larger than in the bulk, evidencing an exceptionally fast electro-optical functionality of the liquid-crystalline-AAO nanohybrids.}

\end{tabular}

 \end{@twocolumnfalse} \vspace{1.6cm}

  ]

\renewcommand*\rmdefault{bch}\normalfont\upshape
\rmfamily
\section*{}
\vspace{-1cm}


\footnotetext{\textit{$^{a}$~Institute of Materials Physics and Technology, Hamburg University of Technology, 21073 Hamburg, Germany. Fax: +49 40 42878-4070 ; Tel: +49 40 42878-3135 ; E-mail: patrick.huber@tuhh.de}}
\footnotetext{\textit{$^{b}$Faculty of Electrical Engineering, Czestochowa University of Technology, 42-200 Czestochowa, Poland ; E-mail: andriy.kityk@univie.ac.at}}
\footnotetext{\textit{$^{c}$~Military University of Technology, 00-908 Warsaw, Poland}}
\footnotetext{\textit{$^{d}$~Helmholtz-Zentrum Berlin f\"ur Materialien und Energie, 14109 Berlin, Germany}}
\footnotetext{\textit{$^{e}$~Institute for the Chemistry of New Materials, University Osnabr\"uck, 49067 Osnabr\"uck, Germany}}
\footnotetext{\textit{$^{f}$~Institute of Optical and Electronic Materials, Hamburg University of Technology, 21073 Hamburg, Germany}}
\footnotetext{\textit{$^{g}$~Institute of Materials Research, Helmholtz-Zentrum Geesthacht, 21502 Geesthacht, Germany}}

\footnotetext{\dag~Electronic Supplementary Information (ESI) available: See ancillary files.}


\footnotetext{$^{\ast}$~Corresponding authors.}



\section{Introduction}

Liquid crystals (LCs) confined at interfaces exhibit properties which often differ markedly from their bulk behaviour. \cite{Crawford1996,Saliba2013,Ocko1986, Ocko1990,Pandey2015,Nych2017} Upon embedding in cylindrical pores \cite{Grigoriadis2011, Zhang2014, Jeong2015, Gim2016, Dietrich2017} and porous media these interfacial pecularities can entirely dominate the liquid-crystalline behaviour. \cite{Araki2011, Kityk2008, Kralj2012, Cerclier2012, Jeong2012, Lefort2014, Kityk2014a, Ndao2014, Calus2014, Lee2015, Calus2015, Kutnjak2003, Kopitzke2000, Stillings2008, Calus2014a, Calus2015c, Xia2015,Chahine2010a, Kityk2010, Huber2015, Ackerman2016,Ryu2017, Kim2017} For example, the phase transitions of nematics in mesopores are significantly changed. The abrupt, bulk isotropic-nematic transformation of first-order can become continuous and characterised by residual paranematic ordering far above the bulk transition temperature. \cite{Kityk2008} Computer simulations of LCs in thin film and pore geometry \cite{Gruhn1997, Care2005, Binder2008, Ji2009, Pizzirusso2012, Karjalainen2015,Cetinkaya2013, Schulz2014, Schlotthauer2015} confirm pronounced spatial heterogeneities, particularly interface-induced molecular layering and radial gradients, both in the orientational order and in the reorientational dynamics in restricted geometries. These effects are not only of fundamental interest, an understanding of these confinement effects is also pivotal to design nanoscopic functional hybrid structures. \cite{Pathak2017, Saliba2013}

\begin{figure}[t]
  \centering
   \includegraphics[width=0.95\columnwidth]{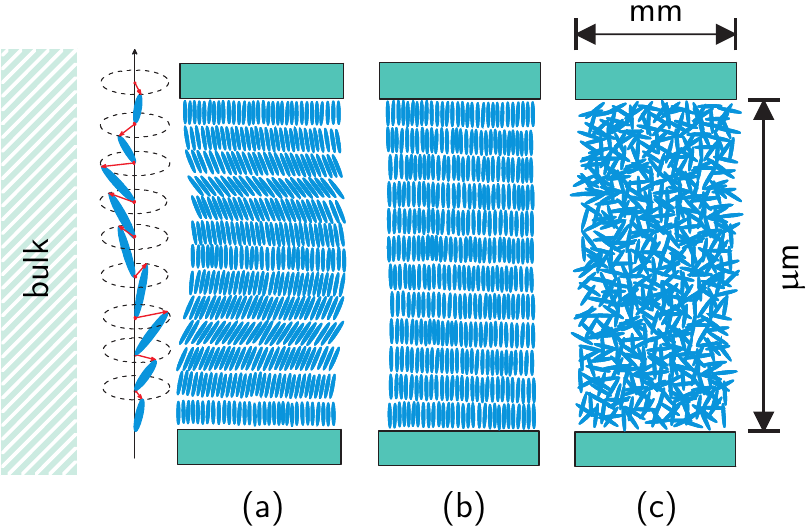}\\
   \vspace{4mm}
   \includegraphics[width=0.95\columnwidth]{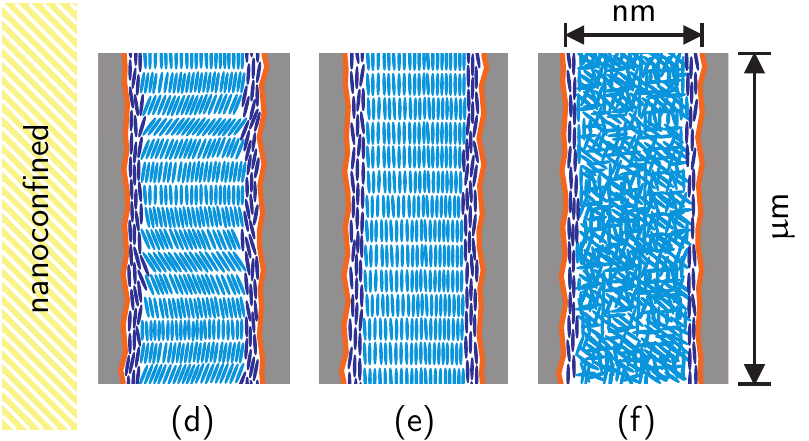}
  \caption{Chiral order in the bulk (homeotropic film geometry): Side-views on the (a) smectic C* phase illustrating the helicoidal changes in the molecular tilt from layer to layer;  (b) smectic A* phase; (c) isotropic liquid phase. Chiral order in nanopores: Side-views on the (d) smectic C* phase; (e) smectic A* phase, (f) isotropic liquid phase. The orange lines indicate the polymeric wall coating facilitating planar mesogen anchoring. The dark blue rods represent mesogens in a paranematic state close to the pore walls.}
  \label{fig:chiral_order_bulk}
\end{figure}

In the past mainly non-chiral mesogens \cite{Als-Nielsen1982, Yokoyama1988, Fukuto2008, Aya2011, Ruths2012} have been explored with respect to these phenomenologies and there are only a few studies regarding the properties of confined chiral LCs. \cite{Fukuda2011, Inoue2011,Melle2014, Schlotthauer2015,Calus2016, Guo2016, Afghah2017} Compared to non-chiral LCs molecular chirality results in additional structural features, in particular in the formation of helical structures.\cite{Kumar2001,Kitzerow2001, Lagerwall2006} Chiral nematics (N*-phases) or so-called cholesterics, exhibit no positional ordering but collective orientational ordering in one direction within planes. The direction changes from plane to plane, \textit{i.e.}\ it rotates around an axis perpendicular to the orientationally ordered layers with a certain periodicity, the cholesteric pitch. In the chiral tilted smectic phases (SmC* phases) of so-called ferroelectric liquid crystals (FLC) molecules are positionally ordered within smectic layers and show a gradual change in the tilt direction from layer to layer, such that the director precesses conically around the helix axis with a certain periodicity, see Fig.~\ref{fig:chiral_order_bulk}a.\cite{Meyer1975} This helical pitch $p$ changes with temperature $T$ and often ranges from several hundreds to several thousands layers, \textit{i.e.}\ it corresponds to wavelengths of ultraviolet, visible or infrared light. 
Resulting from this remarkable phase behaviour chiral LCs exhibit unique optical properties,\cite{Kitzerow2001} such as circular dichroism, large optical activity, Bragg reflection, \cite{Dunmur1999} electro-optical effects \cite{Skarabot1998a,Skarabot1999} and low-threshold laser emission. \cite{Kopp1998} These effects have been exploited extensively in applications employing bulk chiral LCs.\cite{Ha2008, Castles2014, Srivastava2015} 

Another remarkable feature of the SmC* phase is the in-plane net polarization of a single smectic layer induced due to the symmetry breaking of the medium comprising chiral and polar molecules which undergo a hindered rotation.\cite{Meyer1975, Raszewski1995} The polarization vector $\vec{P}$ of each single smectic layer rotates along the helices, thus macroscopically the polarization averages to zero in bulk samples. Upon application of an external electrical field transverse to the helical axis,  the resulting force momentum vector $\vec{M}=\vec{P} \times \vec{E}$, where $\vec{E}$ is the electric field vector, allows one to unwind the helix. In fact, already the interaction between the LC and bounding plates can be sufficient to unwind the intrinsic helical structure by confining an FLC in a thin homogeneous cell (with a gap smaller than the helical pitch).\cite{Clark1980}  Such a surface stabilized ferroelectric liquid crystal arrangement is of great practical importance, since it allows for a fast switching in display and opto-electronic applications. \cite{Kitzerow2001, Srivastava2015}  

Previous studies on the structure and dynamics of confined chiral LCs were mainly performed for pore sizes in the submicron range ($\ge$0.2~\textmu m).  \cite{Schmiedel1994,Binder1996,Rozanski1999,Rozanski2001,Sandhya2002,Rozanski2006,Rozanski2008,Rozanski2011} Whereas the molecular mobility of the single molecules were reported to be unchanged, collective modes were found to be entirely suppressed \cite{Rozanski1999, Rozanski2001,Rozanski2006} or changed in their dynamics. \cite{Sandhya2002, Rozanski2008,Rozanski2011} For cylindrical channels with 0.2~\textmu m diameter an extreme increase in the relaxation frequency of collective helicoidal modes in the SmC* phase was tentatively attributed to a partial unwinding of the helices due to the surface fields at the pore walls. \cite{Sandhya2002} To the best of our knowledge, molecular ordering and mobility of FLCs inside nanopores less than 50~nm has not been studied so far. Here, we report optical birefringence, neutron diffraction, as well as electro-optical experiments on the chiral FLC \mbox{2MBOCBC} embedded into nanopores of anodic aluminium oxide with different radii ranging from about 10 to 20~nm.\cite{Lee2014} We explore whether in such extreme spatial confinement helical SmC* structures can form and which collective dynamical behaviour results.

\section{Experimental}

\begin{figure}[htb]
	\centering
	\includegraphics[angle=0,width=0.92\columnwidth]{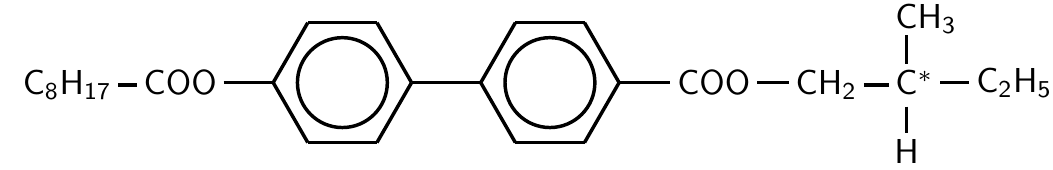}\\
	\vspace*{1mm}
	\includegraphics[angle=0,width=0.92\columnwidth]{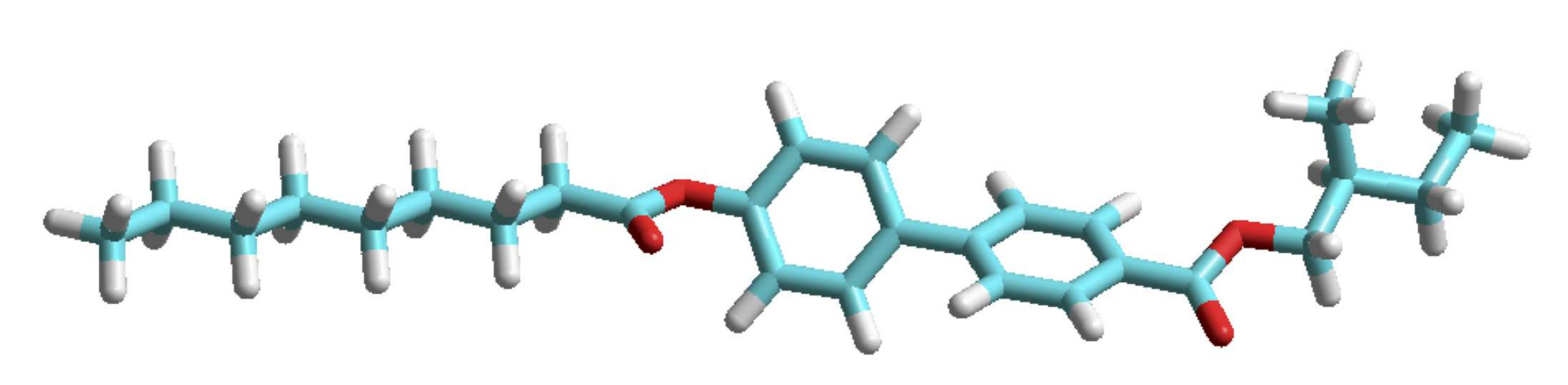}
	\includegraphics[width=0.9\columnwidth]{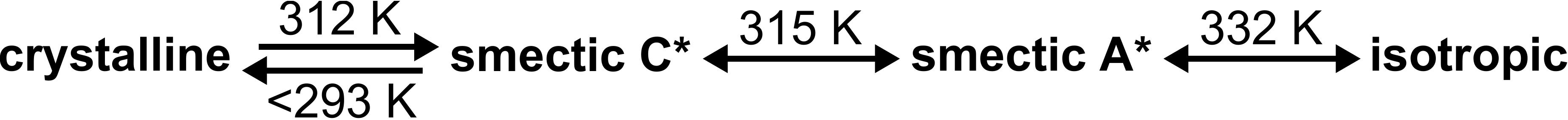}\\
	\vspace*{3mm}
	\includegraphics[width=0.45\columnwidth]{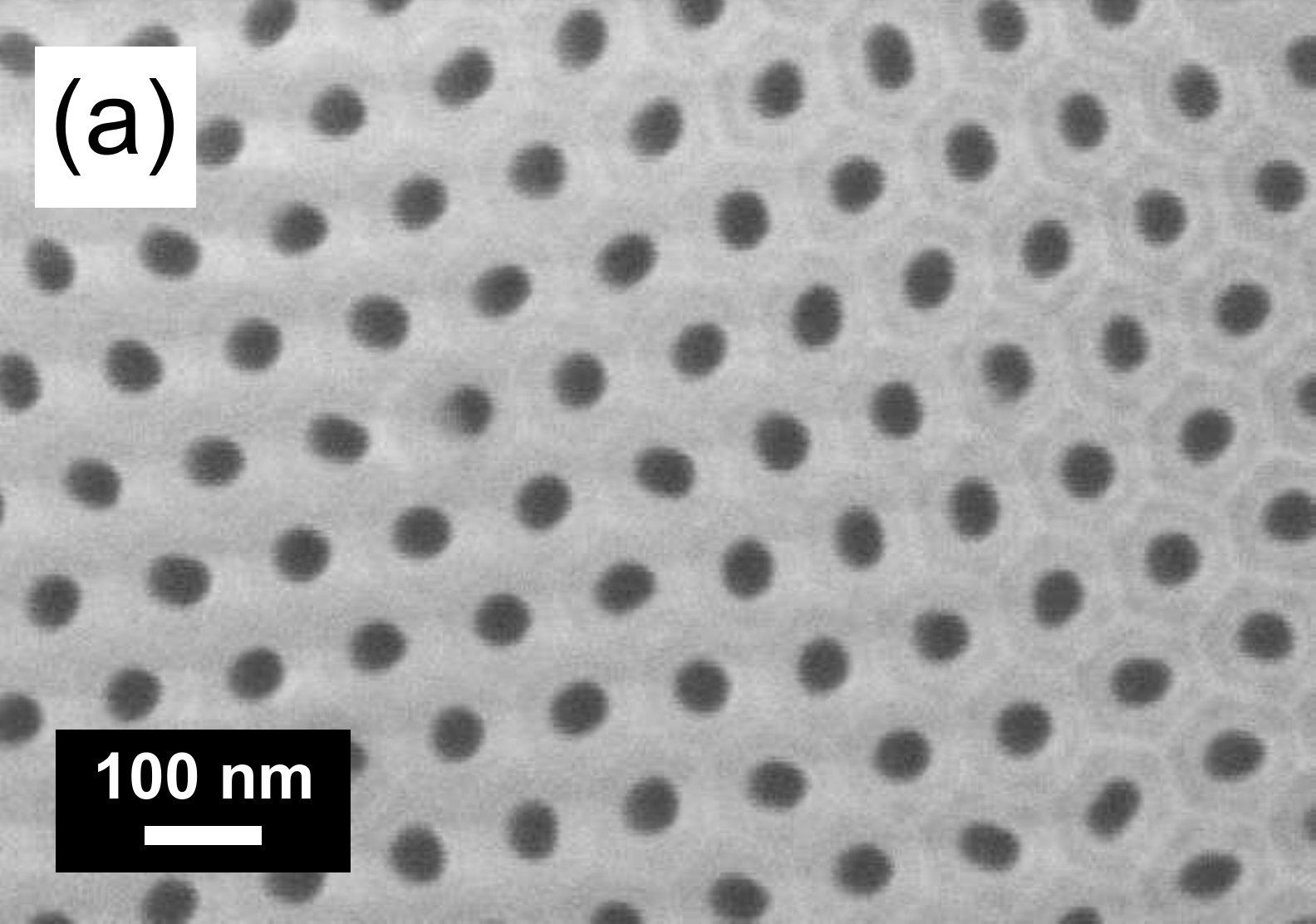}
	\hspace{2mm}
	\includegraphics[width=0.45\columnwidth]{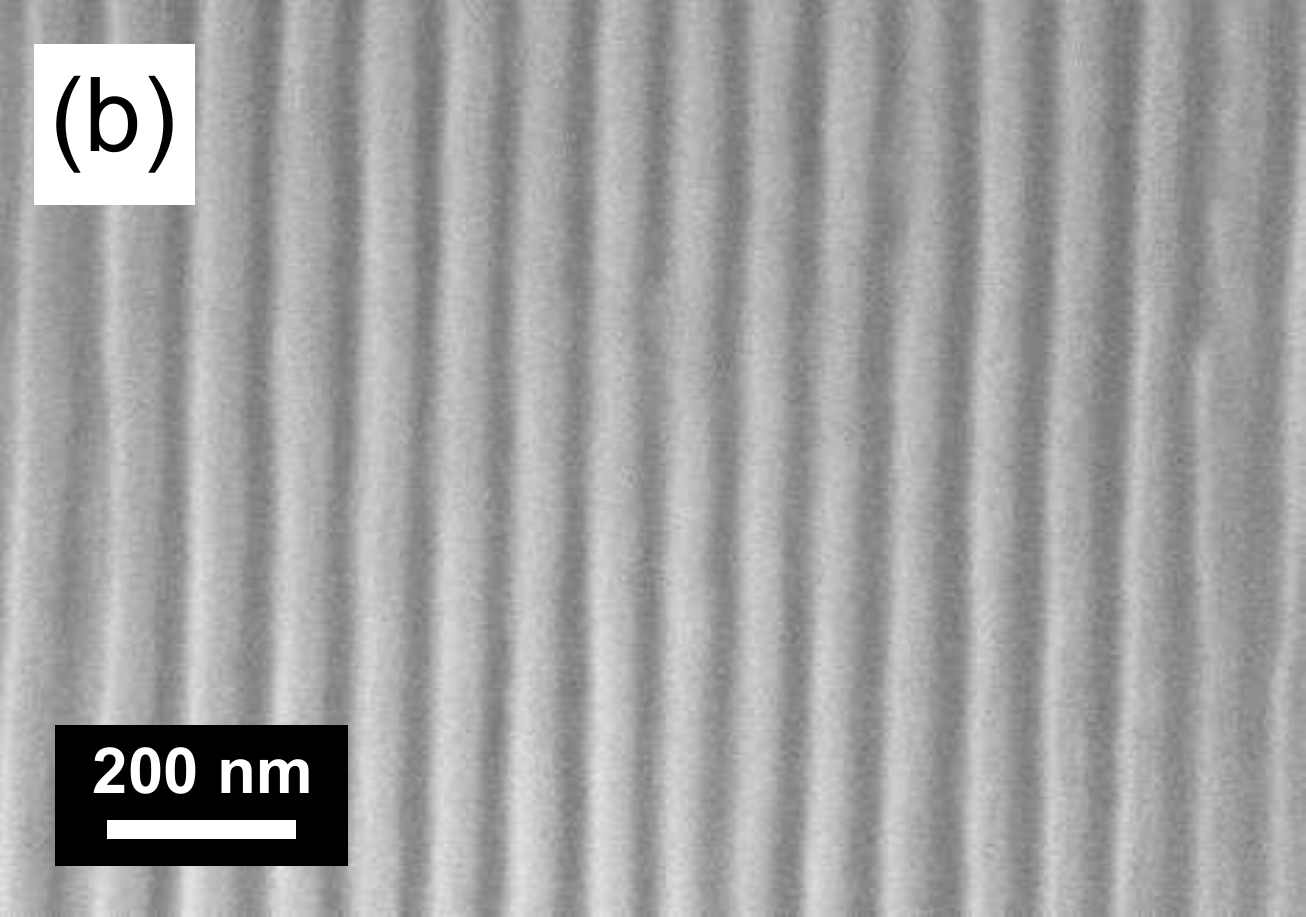}
\caption{Chemical structure of \mbox{2MBOCBC} along with the characteristic phases and bulk phase transition temperatures. Scanning electron microscopy top (a) and side (b) views on a mesoporous anodic aluminium oxide membrane with a pore radius of 21~nm.}
\label{fig:chemical_structure}
\label{fig:SEM_Al2O3}
\end{figure}

The chemical structure of the FLC S-(-)-2-methylbutyl 4-n-nonanoyl\-oxy\-bi\-phenyl- 4$'$-carboxylate (hereafter denoted \mbox{2MBOCBC}, known also as IS-2424) is depicted in Fig.~\ref{fig:chemical_structure}.
It was purchased from AWAT Ltd.\ (Warszawa, Poland). In the bulk state this FLC exhibits a sequence of three phase transitions during cooling: It transforms from the isotropic (I) phase to the SmA* phase at 332.6~K, to the SmC* phase at 315.6~K and to the solid crystalline (Cr) phase at 312~K. \cite{Musevic1989,Jang2001,Rozanski2005,Rozanski2005a,Nakao1987,Raszewski1999} The transition to the Cr-phase can be easily supercooled. Thus, the ferroelectric SmC* phase can be observed in a broad temperature range, down to 265~K. \cite{Rozanski2005,Rozanski2005a} In the SmC* phase \mbox{2MBOCBC} is characterised by a helical pitch which decreases with decreasing temperature from about 3.8~{\textmu m} near the SmA*-SmC* transition to about 1.1~{\textmu m} at 281~K. \cite{Musevic1989} Moreover, it exhibits polarization reversal at about 292~K, with the polarization ranging from -2 to +2~nC/cm$^2$. \cite{Nakao1987}

We employ membranes of self-organized anodic aluminium oxide (AAO) consisting of a parallel array of nanopores of different diameters aligned perpendicular to the membrane surface. The AAO membranes (thickness $h$ =100~{\textmu m}), synthesized in-house as well as bought from Smart Membranes GmbH (Halle, Germany), were fabricated by electrochemical etching from pure aluminium. 
The pore radii $R$ and porosities $P$ are determined by volumetric N$_2$-sorption isotherms at $T=77$~K to 21.0$\pm$2.0~nm ($P=23$~\%), 15.5$\pm$1.5~nm ($P=18$~\%) and 10.0$\pm$0.7~nm ($P=16$~\%), respectively. Scanning electron micrographs of the $21.0$~nm membrane are shown in Fig.~\ref{fig:SEM_Al2O3}.

The native, polar surfaces of AAO membranes often favor tangential anchoring of LCs and thus in nanopores an axial type of ordering results. However, this is not the case for AAO-\mbox{2MBOCBC} nanocomposites where the tangential anchoring appears to be metastable and after several cooling-heating cycles a normal anchoring becomes dominant. The as-prepared nanoporous AAO membranes do not provide a stable and well-defined wall anchoring condition, see the Electronic Supplementary Information (ESI)$^\dag$ for optical birefringence measurements during cooling-heating cycles demonstrating the unstable anchoring conditions. We find that a nanometric film of the polymer SE-130 deposited on the pore walls is able to enhance the tangential anchoring significantly. It results in thermal history-independent optical and diffraction experiments. For the preparation of the polymer coating the procedure described in Ref.~\cite{Calus2016} is followed. Birefringence measurements on a nematic LC in samples with and without coating and nitrogen sorption isotherms on an empty membrane before and after coating the surface indicate an average thickness of the polymer coating of less than 2~nm. Subsequently, the membranes are imbibed with the FLC melt by pore wetting at a temperature a few degrees above the I-SmA* phase transition. \cite{Gruener2011}

Our optical polarization investigations combine two types of measurements performed simultaneously on a sample: i) the optical rotation $\Psi$ is measured along the optical axis of the sample which is parallel to the long axis of the pores and ii) the optical retardation $\Delta$ caused by the optical birefringence is measured at a fixed angle ($\alpha \sim$ 36$^\circ$) of the incident laser light ($\lambda=633$~nm) with respect to the optical axis, see Fig.~\ref{fig:Birefringence}.
The corresponding set-up is based on polarized light modulation employing a photoelastic modulator (PEM) PEM-90 (Hind Instruments), see Fig.~1 in the ESI$^\dag$. The set-up is similar to the one used in Ref.~\cite{Skarabot1998}, however, here only one PEM is used. The spectra of the modulated intensity, detected by photodetectors PD1 and PD2 depend on the sample characteristics $\Psi$ and $\Delta$, respectively, and are analysed for each arm by two lock-in amplifiers that measure the amplitudes of the first ($I_{\Omega}$) and second ($I_{2\Omega}$) harmonics of the modulated light intensity. The optical rotation, $\Psi$ and optical retardation $\Delta$, measured simultaneously in the $\Psi$- and $\Delta$-arm, respectively, are related to the measured amplitude intensities by the following equation:
\begin{equation}
\Psi=\frac{1}{2}\arctan(kI_{\Omega}/I_{2\Omega}),~ \Delta=\arctan(kI_{\Omega}/I_{2\Omega})
\end{equation}
where the effective coefficient $k=r(2\Omega)\cdot J_2(A_0)/(r(\Omega)\cdot J_1(A_0))$, $r(\Omega)$ is the frequency response function of the photodetector, $J_1(A_0)$ and $J_2(A_0)$ are the values of Bessel functions at the amplitude of PEM retardation $A_0$. In our studies, we use a modulation frequency $\Omega/2\pi$ of 50~kHz and the retardation amplitude $A_0$=0.383$\lambda$. The effective coefficient is directly determined within the calibration procedure by performing measurements on etalon samples. The accuracy for both types of measurements is about 0.005$^\circ$.  The sample is placed inside an optical thermostat operated by a temperature controller (Lakeshore-340) with a temperature control accuracy of 0.01~K. The light beams in both arms are arranged in such a way that they probe the identical sample volume.

\begin{figure}[!b]
  \centering
  \includegraphics[angle=0,width=0.9\columnwidth]{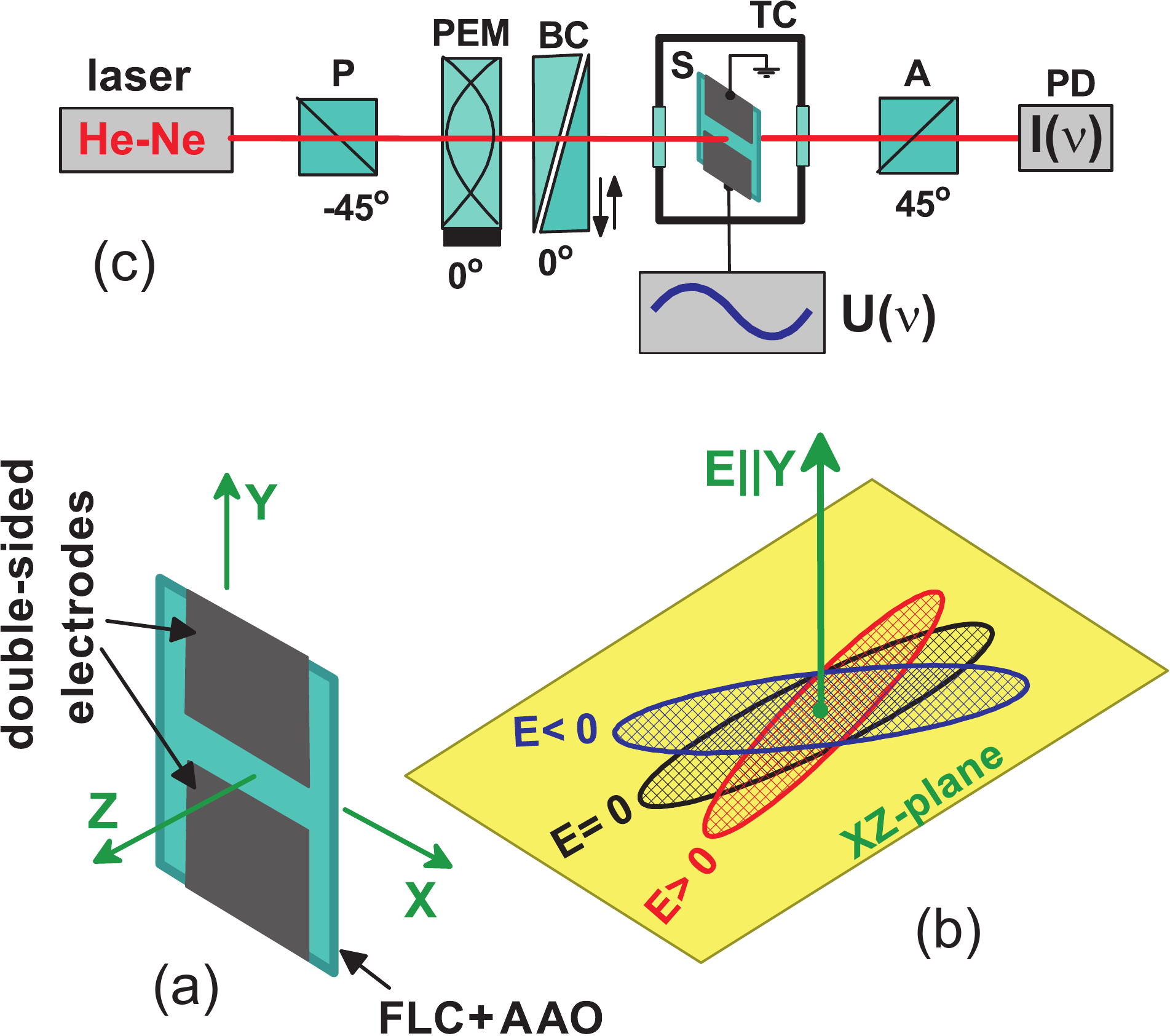}
  \caption{Linear electro-optic response measurements: (a) polarimetry set-up. Here P is the polarizer, A the analyser, PEM the photoelastic modulator, BC the birefringent compensator, and S the sample. The electro-optically modulated light intensity is detected by the photodetector PD and subsequently analysed by a lock-in amplifier. (b) Sample (thin AAO membrane filled with FLC) with attached double-sided aluminium electrodes separated by a gap of $\sim$1 mm. (c) Illustration of the linear electro-optic effect in the nanocomposite or bulk: Applying an electric field perpendicular to the helix axis in the SmC* phase, here along the $Y$-direction, results in the rotation of the optical indicatrix in a plane perpendicular to it, \textit{i.e.}\ in the $XZ$-plane.}
  \label{fig:electro-optic_setup}
\end{figure}

The dynamical properties of the FLC confined in the nanochannels is probed by electro-optical experiments. By means of double-sided surface electrodes separated by a gap of $\sim$1~mm, as sketched in Fig.~\ref{fig:electro-optic_setup}a, an electric field perpendicular to the long axis of the pores of our thin nanoporous membrane (thickness $h=0.1$~mm) can be applied.
Since $a \gg h$ the electric field is only slightly inhomogeneous. The variations are below 10~\% as we estimated by finite element calculations. Because of the alignment of the smectic layers along the nanopores long axes the electric field acts mainly transverse to the helical axes.
A deformation of the helix in an external transverse electric field leads to changes of the refractive indices. Particularly, the $Y$-component of the field induces a change of the dielectric tensor
\begin{equation}
\langle \delta \varepsilon \rangle=\left[ \begin{array}{ccc}
0 \quad& 0 \quad& \langle \delta P_y \rangle \\
0 \quad& 0 \quad& 0 \\
\langle \delta P_y \rangle \quad& 0 \quad& 0 \end{array} \right]
\end{equation}
which is linear in the electric field. It results in a rotation of the optical indicatrix in the $XZ$-plane, which is perpendicular to the electric field vector, see sketch in Fig.~\ref{fig:electro-optic_setup}b. Such a rotation of the optical indicatrix can be detected with a suitable optical technique, see Fig.~\ref{fig:electro-optic_setup}c. To observe the corresponding linear response of the modulated light intensity, which is detected by a photodetector (PD) and measured by a lock-in amplifier, the nanoporous membrane has to be turned out of the optical axis by a rotation (here $\sim$35-45~deg) around the Y-axis. Moreover, the measured electro-optical response depends on the optical retardations induced by the sample and the birefringence compensator. The maximal effect is observable, when their superposition equals $(2n+1)\pi/4$. This can be adjusted for each temperature by the birefringent compensator, BC.

In the case of the bulk measurements a 5~{\textmu m} cell with ITO electrodes and homogeneous alignment is used. The cell is set normal to the laser beam and rotated out of the rubbing direction determining the anchoring orientation of the FLC molecules by 22~deg with respect to the polarization direction of the incident light. The birefringence compensator is used to set the electro-optical response to its maximum level. The measurements are performed with the lock-in amplifier SR-830 having an upper frequency limit of 100~kHz. The measured electro-optical response is normalised by the transmitted light intensity. We present its real and imaginary parts, $\chi'$ and $\chi''$, respectively. Note that this linear electro-optical experiment can be considered as an optical analogue to dielectric relaxation spectroscopy. \cite{Skarabot1998a, Skarabot1999}

Neutron diffraction experiments with a wavelength of 4.567~\AA{} are carried out at the V1 membrane diffractometer of the research reactor \mbox{BER II} of the Helmholtz-Zentrum Berlin. The center of a 128$\times$128 pixel area detector is placed at the Bragg angle of the (001) smectic Bragg peak at $2\Theta=9.4^\circ$, while a stack of identical, surface coated, \mbox{2MBOCBC} filled  membranes is rotated around the $\tilde{\omega}$-axis, perpendicular to the nanopore axis and the incident beam (see sketch of the scattering geometry in Fig.~\ref{fig:neutron_diffraction}a). These rocking scans are repeated for different temperatures during cooling and heating cycles starting and ending in the isotropic phase of the FLC.

\section{Results \& discussion}

\subsection{Thermotropic phase behaviour: Optical birefringence and neutron diffraction}
A preferential orientational ordering of guest molecules in an array of parallel pores results in an excess birefringence, $\Delta n^+$, or associated with this an excess retardation, $\Delta^+ \propto \Delta n^+$. Accordingly, the molecular ordering inside the nanoporous membrane can be precisely characterised by optical polarimetry. In the case of LCs the orientational order is described by the scalar Hermans-Tsvetkov orientational order parameter $S=\frac{1}{2}\langle 3\cos^2\phi-1 \rangle$, where $\phi$ is the angle between a characteristic axis of the molecules and the direction of preferred local molecular orientation (the so-called director $\hat{n}$). The brackets denote here an averaging over all molecules under consideration, whereas the orientation of the director may vary locally depending on anchoring conditions and/or the specifics of the geometrical constraints. Since the wavelength of light is much larger than the pore size the measured retardation scales linearly with $\langle S \rangle_{pv}$, \textit{i.e.}\ the order parameter averaged over the pore volume. 

For optically positive rod-like LCs, as studied here, the axial molecular ordering inside nanopores, originating from a tangential anchoring, results in a positive linear birefringence (or positive retardation). In contrast, a radial or so-called polar molecular arrangement inside the nanopores, resulting from normal anchoring, results in a negative linear birefringence (or negative retardation) with an absolute value approximately two times smaller than in the case of an axial arrangement. Examples of these two types of orderings have been reported in Ref.~\cite{Chahine2010}. The retardation sign can be easily determined in our experiments by a consideration of the phases of the harmonics detected by the lock-in amplifiers, see experimental section.

Molecular chirality brings new features to the optical properties among which \emph{circular} birefringence, also known as optical activity or optical rotation, is a frequently measured quantity used for the characterization of helical structures in cholesteric or SmC* bulk phases. Already in the isotropic phase the chirality of the single molecule leads to an intrinsic molecular optical activity, on the order of a few deg/cm. \cite{Calus2016} However, it is orders of magnitudes larger, about 10$^4$ deg/cm, in the cholesteric or SmC* phase, where collective helical and thus supermolecular chiral structures exist. \cite{Dunmur1999, Kitzerow2001}

\begin{figure}[!htb]
	\centering
	\includegraphics[width=0.75\columnwidth]{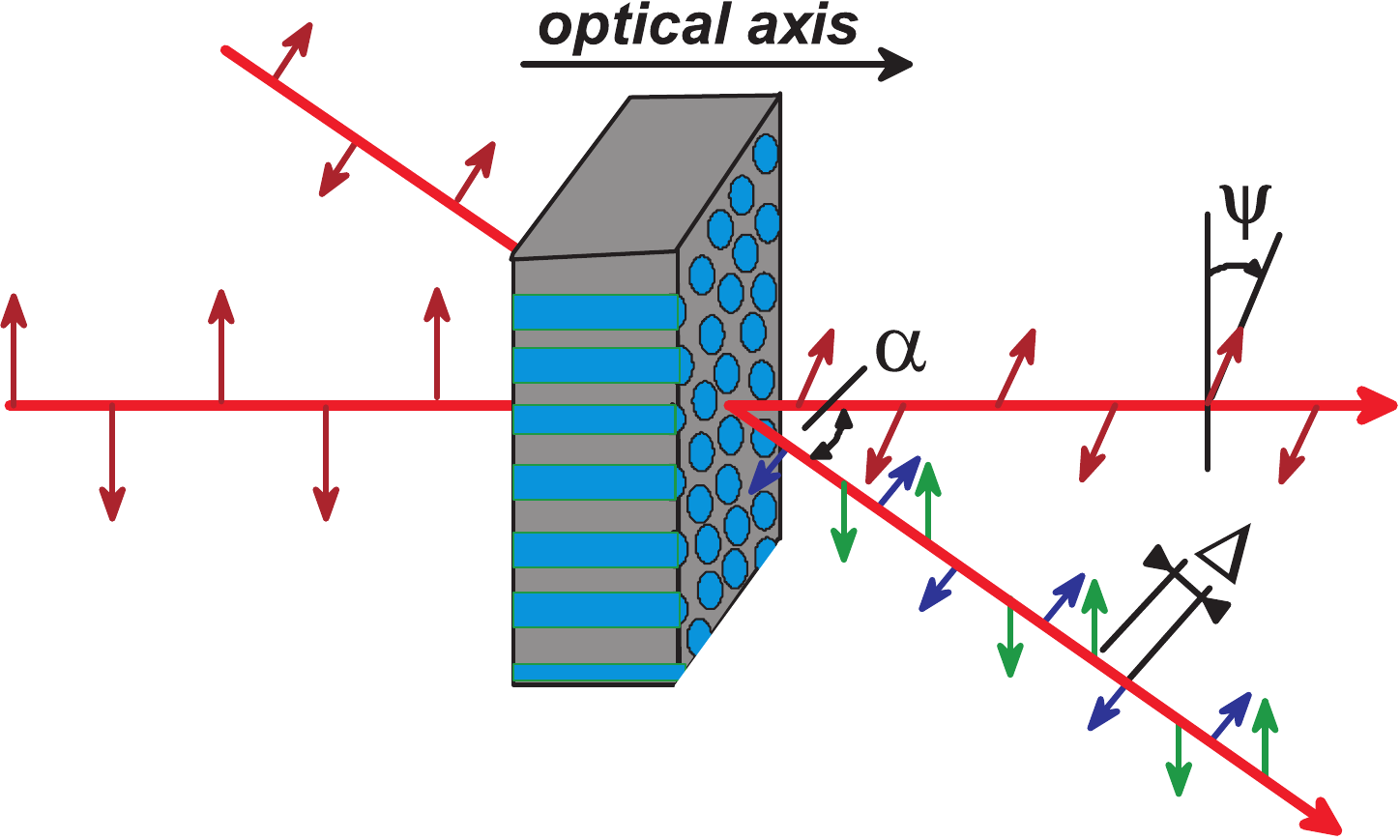}\\
	\vspace{3mm}
	\includegraphics[width=0.95\columnwidth]{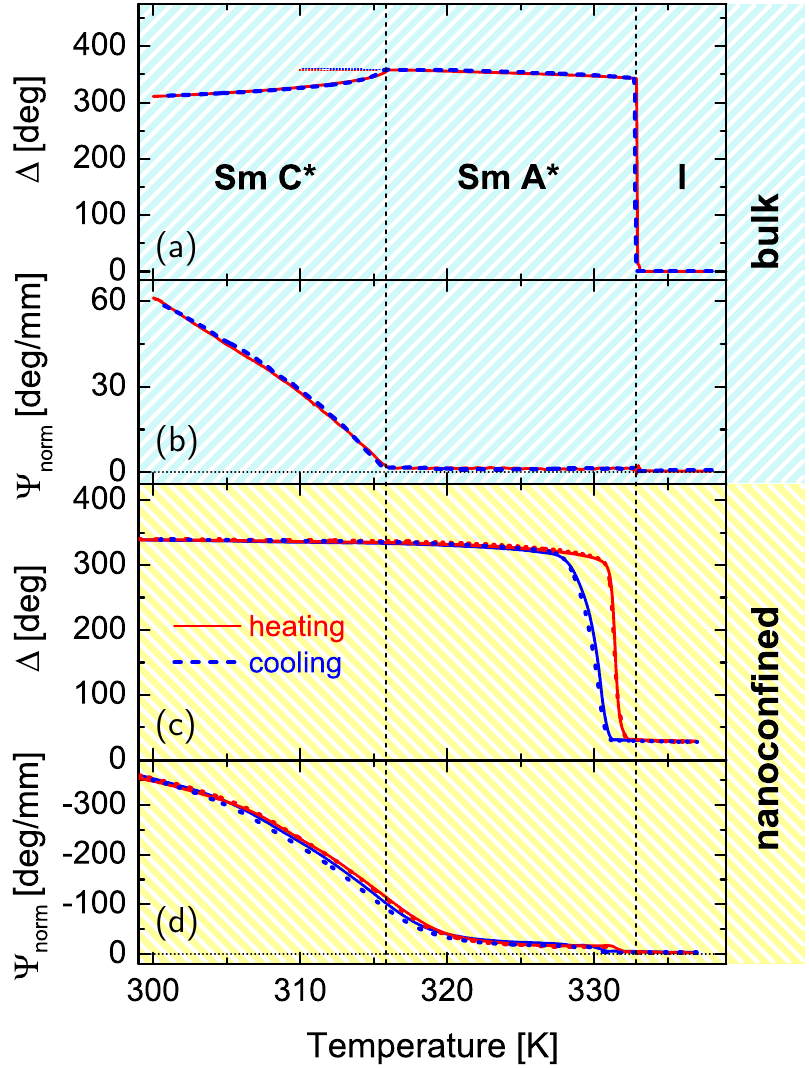}
	\caption{Temperature dependences of the optical retardation, $\Delta$ (a), (c) and normalized rotatory power, $\Psi_{\rm norm}$ (b), (d) of the bulk and confined FLC, respectively. The optical retardation is measured at a fixed angle $\alpha \sim 36^\circ$ as visualized in the sketch. Blue and red curves correspond to cooling and heating runs, respectively.}
	\label{fig:Birefringence}
\end{figure}

Polarimetric measurements on the bulk FLC (30~{\textmu m}-cell, homeotropic alignment) are presented in Fig.~\ref{fig:Birefringence}a and b as a reference for the measurements on the confined system.

\begin{figure}[htb]
  \centering
  \includegraphics[angle=0,width=0.95\columnwidth]{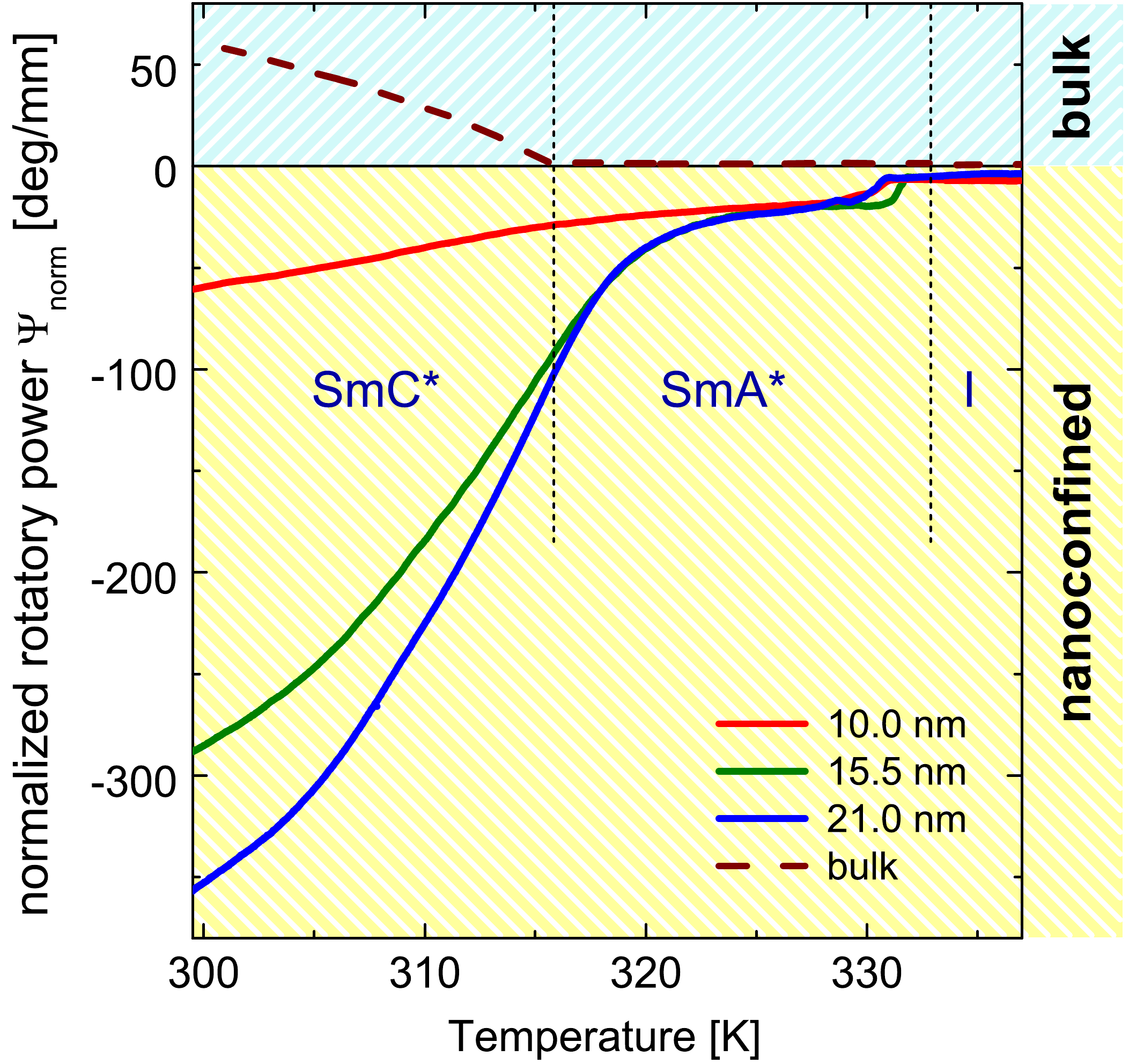}
  \caption{Normalized rotatory power in the bulk state (broken line) and in the confined state for pore radii of $R=21.0$~nm (blue), $R=15.5$~nm (green) and $R=10.0$~nm (red), respectively.}
  \label{fig:NRP}
\end{figure}

The phase transition I-SmA* at $T=T_{IA}^c$ is accompanied by a sudden appearance of an optical retardation and a positive birefringence. It originates from an abrupt collective molecular reorganization from the totally disordered bulk state (Fig.~\ref{fig:chiral_order_bulk}c) to a homeotropic alignment of the molecules with smectic layers parallel to the confining glass plates of the cell (Fig.~\ref{fig:chiral_order_bulk}b), as typical of the first-order character of this phase transition. The kink-like variation of $\Delta n^+(T)$ at $T=T_{AC}^c$ indicates a second order SmA*-SmC* transition. The gradual decrease of the retardation is caused by the tilt of the molecules in the smectic layers and the continuous appearance of the helical supermolecular structure in the SmC* phase. The helix formation, which develops along the direction perpendicular to the smectic layers, is even more evident in the appearance of a large optical activity, see Fig.~\ref{fig:Birefringence}b and \ref{fig:NRP}, below $T_{AC}^c$.

The SmA*-SmC* transition is described by a Landau-de Gennes free energy with even parity terms only. \cite{Levstik1987, Kitzerow2001} The Lifshitz invariant allowed by this symmetry explains the appearing of the helicoidal phase (a long-range modulated structure). Depending on the magnitude of the free energy terms, it can be either of second \cite{Levstik1987, Kitzerow2001} or of first order. \cite{Ratna1988,Soltani2014,SimeaoCarvalho1996} In agreement with our results, a previous study showed that 2MBOCBC exhibits a continuous SmA*-SmC* transition. \cite{Musevic1989} However, the system is close to the tricritical point separating continuous from discontinuous behavior and the critical exponent of the order parameter $\beta$ is 0.26. \cite{Musevic1989} 

The tilt angle $\vartheta$ between the long axis of the molecules and the layering direction plays a decisive role for the behaviour of the optical activity, i.e. it is $\propto \vartheta^4$. \cite{Musevic1989} The tilt angle monotonously increases in the SmC* phase of \mbox{2MBOCBC} up to $\vartheta = 15^\circ$. However, the overall dependence of the optical activity on the geometrical details of the SmC* phase is more complex. \cite{deVries1951} Nevertheless, the optical activity effect in the SmC* phase is at least two orders of magnitude larger than in the SmA* or I phase indicating a collective helical arrangement of the chiral molecules.

The results of polarimetric measurements on the FLC confined in pores with $R=21.0$~nm in direct comparison with the bulk system are shown in Fig.~\ref{fig:Birefringence}. Complementary data sets for the other pore diameters studied show a similar temperature behaviour and can be found in the supplement. All nanocomposites exhibit a saturated positive birefringence at low temperature, which decreases upon heating to a small positive value at about 331-332~K, \textit{i.e.}\ close to the temperature of the bulk clearing point. This indicates a transition from a state with collective arrangement of the molecular long axes parallel to the long pore axis at low temperature to an isotropic orientational order at high temperature. The small positive optical retardation in the isotropic state results from geometric birefringence \cite{Kityk2009, Calus2014} caused by a difference in the isotropic refractive indices of the host matrix and guest LC materials in combination with the elongated nanopore geometry. Moreover, the mesogens in direct proximity of the pore walls, which are known to be in a paranematic state due to the anchoring field at the pore walls, \cite{Sheng1976, Kityk2008} contribute to this high temperature birefringence, \cite{Kityk2008, Calus2012, Huber2013} see the dark blue highlighted mesogens in Fig. \ref{fig:chiral_order_bulk}d, e, f.

There is a sizeable cooling-heating hysteresis at the I-N transition in the confined compared to the bulk state. It is well known that the nucleation of the low- and high-temperature phase can occur via quite different thermodynamic paths and at different locations (e.g. in the pore center or at the pore walls) upon spatial restrictions. This has been demonstrated for a variety of first-order transitions, most prominently for the crystallization of simple \cite{Huber1999, Christenson2001, Alba-Simionesco2006,Khokhlov2007,Schaefer2008, Knorr2008, Schappert2013} and complex basic building blocks. \cite{Morishige1999, Duran2011,Kityk2014a, Soprunyuk2016} For example, we inferred that for certain textures of discotic liquid crystals the low-temperature phase is nucleated at the pore wall and propagates than towards the pore center upon cooling and vice versa upon heating. \cite{Kityk2014a} Detailed additional experimental studies, in particular as a function of pore filling, could possibly clarify the different nucleation paths. \cite{Huber2013} We suggest the following scenario: The absence of free surfaces and the small volumes contained in single nanopores could result in the absence of heterogeneous smectic nuclei and thus result in crystallization at the low, homogeneous phase transition temperature. Upon heating the aligned smectic phase is stabilized by the cylindrical pore geometry. Thus, it melts at a high temperature, which is however still lower than the bulk melting temperature, since the confined ordered phase has to accommodate to the extreme spatial confinement. \cite{Huber1999}  

\begin{figure}[htb]
  \centering
  \includegraphics[width=0.95\columnwidth]{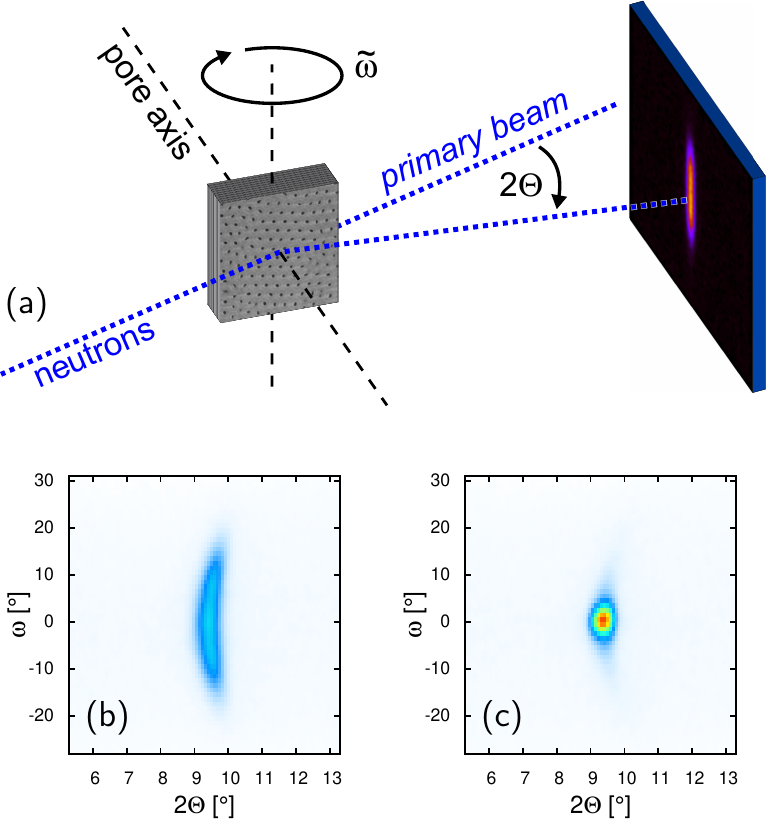}
  \caption{(a) Scattering geometry of the neutron diffraction experiment and intensity contour maps of the smectic diffraction peak as a function of sample rotation $\tilde{\omega} = \omega + \Theta$ and scattering angle 2$\Theta$ at $T=279$~K (b) and $T=$324~K (c), respectively.}
  \label{fig:neutron_diffraction}
\end{figure}

\begin{figure}[htbp]
	\centering
	\includegraphics[width=0.88\columnwidth]{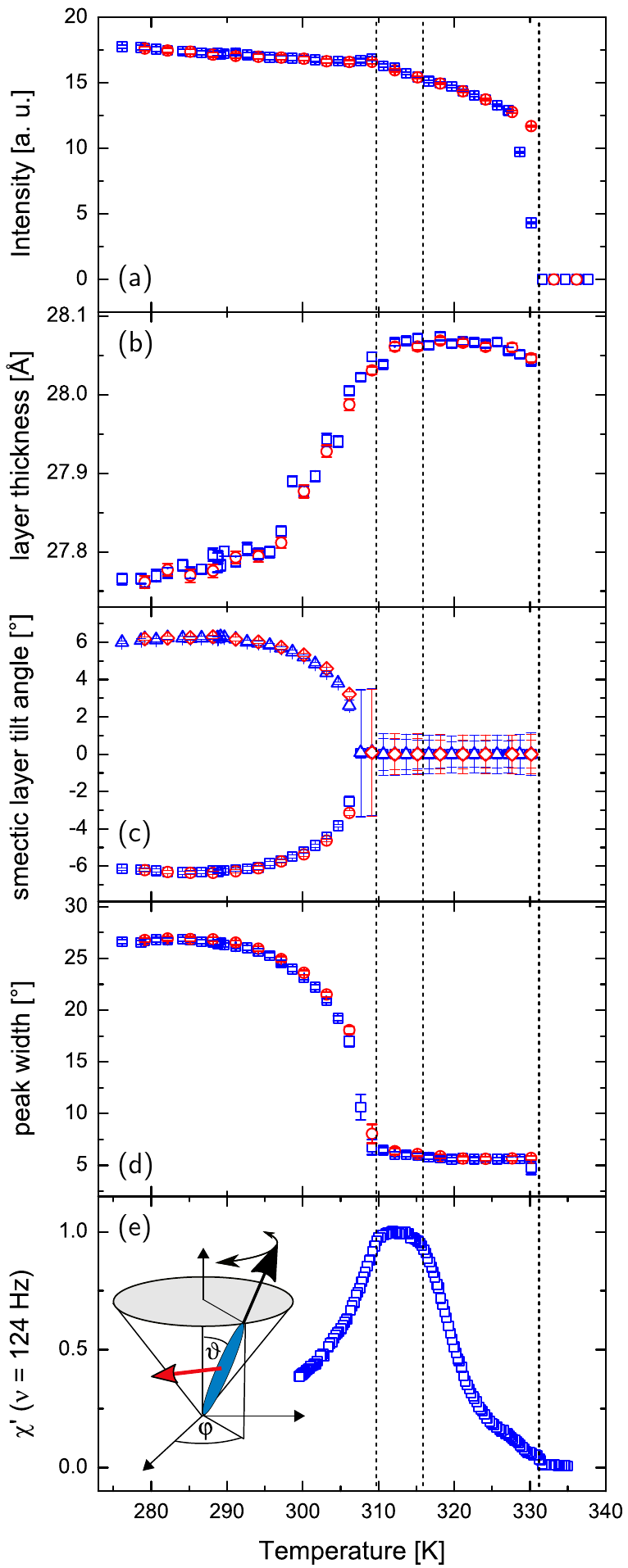}
	\caption{Neutron diffraction and electro-optical experiments on the FLC confined in 21~nm nanopores: (a) Integrated neutron scattering intensity of the (001) smectic layering peak, (b) thickness of the smectic layers, (c) tilt angle of the smectic layer planes, (d) peak width with respect to $\omega$ rocking scans, (e) real part of the normalized linear electro-optical response, measured at a constant frequency of 124~Hz corresponding to the quasi-static regime. The inset illustrates the orientation of the director within each smectic layer as parametrized by the phase angle $\phi$ and tilt angle $\vartheta$ as well as the polarization vector $\vec{P}$ of a single smectic C* layer (red arrow). Blue squares correspond to the cooling and red circles to the heating cycle, respectively.}
	\label{fig:Neutron_Intensity_layer-thickness_peak-width_electro-optic-response}
\end{figure}

\begin{figure}[ht]
	\centering
	\includegraphics[width=0.5\columnwidth]{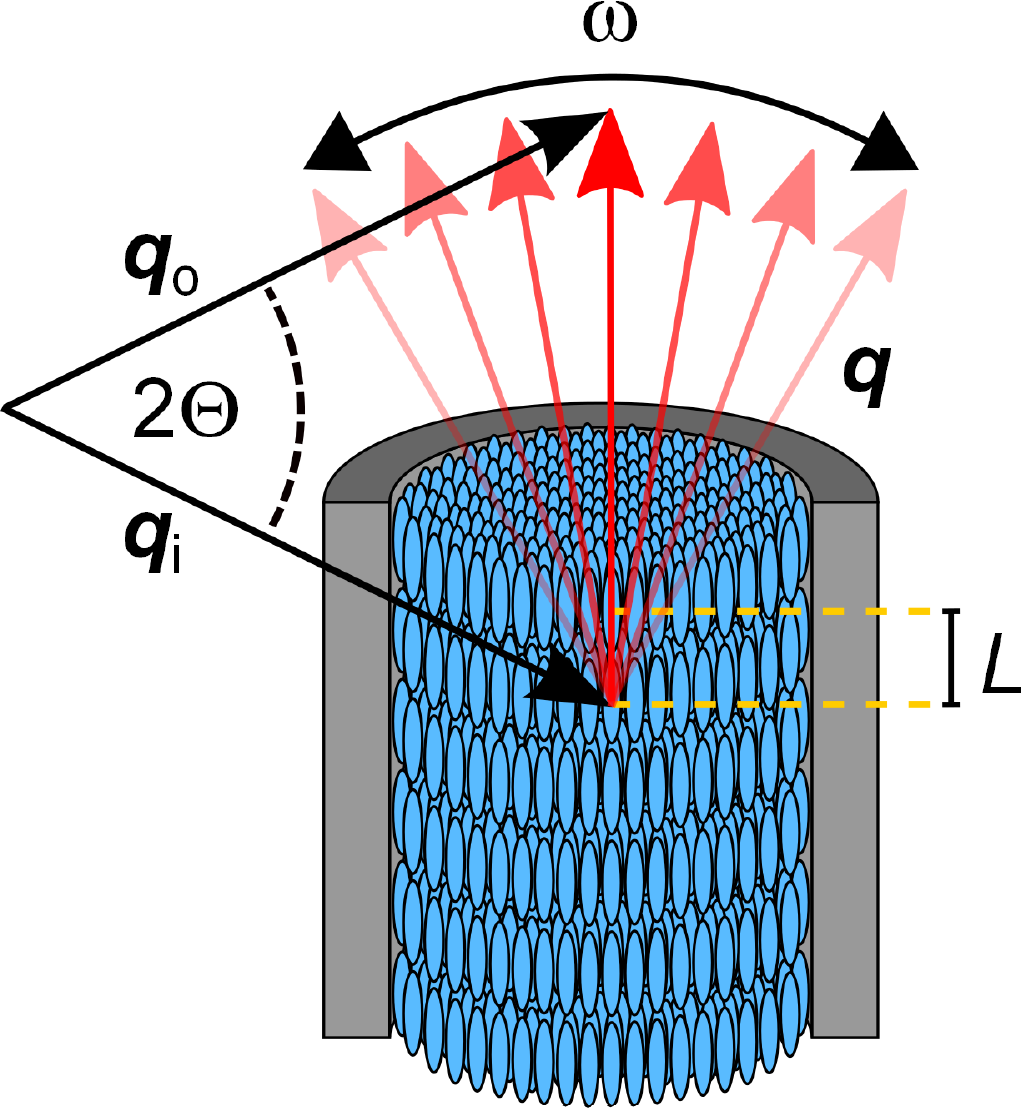}\\
	\vspace{5mm}
	\includegraphics[width=0.95\columnwidth]{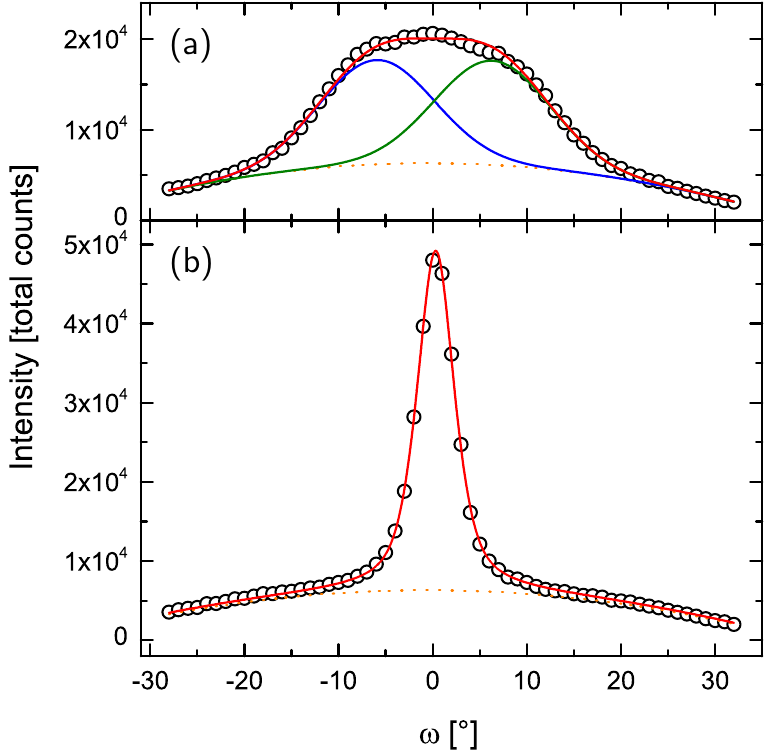}
	\caption{Geometry of the neutron scattering experiment and integrated intensity of the $\omega$-rocking scans, as depicted in Fig.~\ref{fig:neutron_diffraction}, for (a) $T=279$~K and (b) $T=324$~K. Red lines are the total fit functions, blue and green lines are the single peaks (composing to the total fit function), already containing the fixed background, plotted as dotted curve.}
	\label{fig:omega-scan-fits}
\end{figure}

Neutron diffraction experiments probing the translational FLC order in the pores are consistent with the findings in the optical experiments. In Fig.~\ref{fig:neutron_diffraction} the intensity of the Bragg peak typical of the smectic layering of the confined FLC is shown as a function of the sample rotation $\tilde{\omega} = \omega + \Theta$ and scattering angle $2\Theta$, as indicated in the scattering geometry (a), for $T=324$~K (c) and $T=279$~K (b), respectively.
The integrated intensity of this peak steeply rises from zero at $T=T^c_{IA}$ when cooled from the isotropic phase (see Fig.~\ref{fig:Neutron_Intensity_layer-thickness_peak-width_electro-optic-response}a), analogous to the increase of the optical birefringence (see Fig.~\ref{fig:Birefringence}c). Moreover, the (001) Bragg reflection is sharply peaked at $\omega=0^\circ$ and thus indicates an arrangement of the layering direction parallel to the long axes of the nanopores (see sketch of the scattering geometry in Fig.~\ref{fig:omega-scan-fits}).

The temperature-dependent normalised rotatory power is shown in Fig.~\ref{fig:Birefringence}d. It is the optical rotation normalised to the thickness and additionally to the porosity of the sample, accounting for the reduced amount of optical active material in the light path compared to the bulk sample (see Fig.~\ref{fig:Birefringence}b). At high temperature it is barely beyond our experimental sensitivity, upon layering formation at $T=T^c_{IA}$ it has a finite value and it is left-handed, \textit{i.e.}\ it is opposite in sign to that observed in the bulk SmA* phase. The rotation power increases continuously with a kink at $T_{AC}^c\sim$~320~K upon cooling and reaches large values indicating the presence of SmC* structures with helical variation of the molecular tilt orientation along the long nanopore axes, see Fig.~\ref{fig:chiral_order_bulk}d. This is corroborated by the temperature evolution of the layering distance as determined from the position of the (001) Bragg peak. It decreases indicating an increasing molecular tilt within the smectic layers as typical of the SmC* phase, see Fig. \ref{fig:Neutron_Intensity_layer-thickness_peak-width_electro-optic-response}b.

The behaviour of the optical activity in the vicinity of the transition from the SmC*-to-SmA* phase likewise in the entire region of the confined SmA* phase requires further discussion. Whereas in the tilted SmC* phase the optical activity is described by the de-Vries equation, \cite{deVries1951, Parodi1975} which gives a large value caused by the helix structure, in the bulk SmA* phase the optical activity is small. It originates solely from the chirality of the single molecules, not from a supermolecular helical structure. In the case of the intrinsic optical activity its value is proportional to the molecular density of chiral molecules. Thus in the composites it is expected to be proportional to the porosity of the nanoporous membranes. In Fig.~\ref{fig:NRP} we compare the rotatory power normalised to porosity (NRP) measured for the three membrane types studied with the one of the homeotropically aligned bulk phase. In the bulk (dashed line) and in confinement (solid lines) they differ in sign as well as in their absolute value. Remarkably, in the confined state the normalised optical activity is by more than one order of magnitude larger than in the bulk state for the SmA* phase and at lower temperature still by up to a factor of six enhanced compared to the bulk SmC* phase (for the 21~nm channels) and it increases significantly with $R$ at lower temperatures.
 
What are the reasons for such differences between the bulk and the confined state? Whereas the sign change in the optical rotation is understandable by a change in the pitch length upon confinement compared to the wavelength of our laser light, \cite{deVries1951, Parodi1975} a large optical activity in the confined SmA* phase indicates a pretransitional supermolecular chiral structure. Presumably smectic layers with tilted molecules, exist in the nanopores. Either they are located in the interfacial layers (close to the pore walls) and result from the bonding of molecules to the polymer interface layer under a certain angle (a pretilt angle) which depends on the specifics of the host/guest interaction and can be a few degrees. \cite{Rasing2004} Or the classic confined SmA* phase is suppressed even in the pore center in favor of a parasmectic C* phase, with already slightly tilted molecules due to the interactions with the interfacial layers, which themselves are often rather in a non-layered, paranematic state due to geometrical and chemical wall heterogeneities and the channel wall anchoring field, \cite{Kityk2010, Calus2012} see Fig.~\ref{fig:chiral_order_bulk}d,~e.

The existence of this pretransitional effects would explain the continuous evolution of the optical activity and birefringence at the SmC*$\rightleftharpoons$SmA* transition. It is also compatible with the thermal evolution of the smectic layer thickness $d_A$ inferred from the neutron scattering experiments, see Fig.~\ref{fig:Neutron_Intensity_layer-thickness_peak-width_electro-optic-response}b. More importantly, the increase in the normalised rotatory power with increasing channel diameter in the SmC* phase, see Fig.~\ref{fig:NRP} corroborates the assumption that the enhanced optical activity originates in the core volumes rather than in interfacial layers. This volume increases with the square of the channel radius $R$, whereas the contribution of the interfacial layers decreases with $1/R^2$. However, a full quantitative analysis of this effect in the SmC* phase is difficult, since a change in the pitch length may also contribute to the variation in the NRP as a function of $R$. Thus, a full understanding of the behaviour necessitates studies of radial structural gradients, for example by filling-fraction-dependent experiments, \cite{Huber2013} as well as an exploration of the pitch length in the confined state. 

Note that the optical activity in the 10~nm-pores is the smallest in the entire temperature regime studied. Also the optical birefringence and thus the orientational order evolves for this smallest pore diameter much more gradual as a function of temperature, see Fig.~3 in the supplement, indicating that the entire liquid-crystalline order is more significantly disturbed than in the larger pores. Presumably, this results from the significantly stronger geometrical constraints. However, also the contribution of disordered mesogens in direct wall proximity, which results from chemical heterogeneities and wall roughness in nanoporous media \cite{Crawford1996, Kityk2010} is the largest for this pore diameter, see Fig.~\ref{fig:chiral_order_bulk}.

\begin{figure}[t]
	\centering
	\includegraphics[width=0.95\columnwidth]{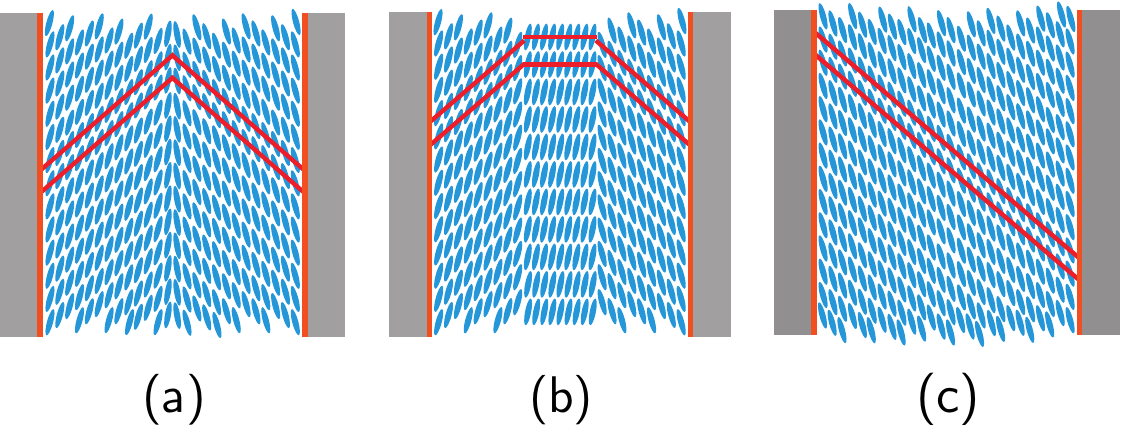}
	\caption{Idealised side-view of (a) chevron-like FLC smectic C* structure, (b) combination of ''normal'' smectic C* and chevron-like FLC structure and (c) smectic layer planes tilted in only one direction inside the nanopores, disregarding the helix structure along the pore axis for visual clarity. It should be kept in mind that in reality the geometry in (a) and (b) exhibits a three-dimensional structure with a rotational symmetry around the long pore axis. The smectic layer planes are indicated by red lines.}
	\label{fig:chevron-structure}
\end{figure}

The temperature-dependent neutron scattering experiments reveal another peculiarity of the molecular arrangement in the pores. After a sharp increase in the smectic layering peak intensity (integrated over $2\Theta$ and $\omega$) at $T=T^c_{IA}$, the intensity slowly increases upon cooling until approx.\ 309.5~K, where a downward kink in the intensity curve occurs (see Fig.~\ref{fig:Neutron_Intensity_layer-thickness_peak-width_electro-optic-response}a). More interestingly the width of rocking scans, \textit{i.e.}\ the width of the smectic layering peak with respect to $\omega$ variations, dramatically increases (see Fig.~\ref{fig:Neutron_Intensity_layer-thickness_peak-width_electro-optic-response}d) at this temperature, compare also the intensity maps of Fig.~\ref{fig:neutron_diffraction}b and c, as well as the video in the ESI$^\dag$. This suggests the formation of layers tilted with respect to the long pore axis direction ($\omega=0^\circ$). An analysis of the rocking peak shape reveals that the high temperature phase can be well described with one Pseudo-Voigt peak, whereas at least two peaks are necessary for a description of the rocking scans at lower temperatures (see Fig.~\ref{fig:omega-scan-fits}).
The temperature evolution of the tilt angle obtained from the position of each peak with respect to $\omega=0^\circ$ is shown in Fig.~\ref{fig:Neutron_Intensity_layer-thickness_peak-width_electro-optic-response}c. Both the tilt angle and the total peak width (Fig.~\ref{fig:Neutron_Intensity_layer-thickness_peak-width_electro-optic-response}d) increase in the same manner and the rocking peak splits symmetrically in two subpeaks. At around $T=291$~K the layering tilt angle saturates at approx.\ $6^\circ$ and stays constant upon further cooling. Upon heating this behaviour is completely reversible.
The single peak, centered at $\omega=0^\circ$, corresponds to smectic layers perpendicular to the pore axis. However, a symmetric splitting into two peaks around $\omega=0^\circ$ indicates the formation of layers equally tilted with respect to the pore long axes, as schematically shown in Fig.~\ref{fig:chevron-structure}. 

There are a number of possible orientational domain structures compatible with the splitting of the rocking scan peak, the most prominent one is the chevron-like structure shown in Fig.~\ref{fig:chevron-structure}a.
Chevron-formation was already reported for LCs confined between glass plates, \textit{i.e.}\ in a slab at least two orders of magnitude larger than the pore diameter studied here.\cite{Pelzl1979,Rieker1987,Ouchi1988,Takanishi1989} Also for 2MBOCBC it has been already observed for such slab geometries.\cite{Jang2001} To maintain the increasing molecular tilt within the SmC* layers and the planar anchoring at the walls the smectic layers start to buckle. \cite{Rieker1987, Jang2001, Lagerwall2006} The molecules are translationally anchored along the confining pore walls, with the period of the smectic A* layer spacing $d_A$. At the transition to the tilted SmC* phase the layers normally start to contract while the molecules at the walls preserve their anchoring. The only possibility compatible with both the wall anchoring with period $d_A$ and the SmC* structure with period $d_C$ with $d_C<d_A$ is that the layers buckle into the chevron geometry. \cite{Lagerwall2006}
The peak at low temperatures can be described by two symmetric peaks with the same intensity. This indicates that the domain size of the two distinct tilt directions averaged over the sample and thus over all AAO pores are identical. However neither a rotational symmetry about the central axis of each pore must exist nor a homogeneous domain structure within each pore. For example, a coexistence of different domain structures has been reported for block copolymers confined in AAO membranes. \cite{Wang2009, Yu2006} Orientational domains with only one tilt direction are also imaginable, see Fig.~\ref{fig:chevron-structure}c. Since the tilt direction of the layers changes randomly from pore to pore or even from orientational domain to orientational domain within one single pore, the resulting averaged diffraction signal would agree with the one typical of the symmetric chevron structure. However, a collective tilt of smectic layers in one direction over the whole pore diameter would require a coherent shift of the molecules in axial direction, in particular at the pore walls. The translational rearrangements necessary for the formation of the chevron-like structures with revolutionary symmetry around the long pore axis from the bookshelf arrangements are thus much more likely. In fact, this kind of smectic layer buckling was recently also reported for an achiral SmC phase confined in AAO nanopores. \cite{Lefort2014}  

Note also that the chevron-formation exhibits a well defined temperature onset as well as a remarkable thermal cycling robustness, as can be inferred from the complete reversibility between cooling and heating scans in all diffraction parameters in Fig.~\ref{fig:Neutron_Intensity_layer-thickness_peak-width_electro-optic-response}. In the following, we will see that this structural rearrangement affects also the collective molecular dynamics and thus the electro-optical properties of the nanoconfined FLC.

\subsection{Smectic C* helix dynamics: electro-optical experiments}

In the SmA* phase the molecules are rotating freely around their long axes, so that the molecular director is $\hat{n}= (0,0,n_z)$, where $z$ is the layering direction. Below $T_{AC}$ in the SmC* phase the molecular director tilts away from the normal to the smectic layers. The order parameter of the transition is a two-component tilt vector, $\xi=(\xi_1, \xi_2)=(n_x n_z, n_y n_z)$. The direction of the tilt precesses on the surface of a cone as one goes from one smectic layer to another, thus forming a helicoidal modulation wave:
\begin{eqnarray}
\xi_1 \sim \vartheta \cos{(q_0 z+\phi)} \\
\xi_2 \sim \vartheta \sin{(q_0 z+\phi)}
\end{eqnarray}

Here the tilt angle $\vartheta$ describes the amplitude, $\phi$ the phase of the modulation wave and $q_0=2 \pi/p$ quantifies the pitch of the helix, see inset in Fig.~\ref{fig:Neutron_Intensity_layer-thickness_peak-width_electro-optic-response}e. There are two characteristic order-parameter excitations.\cite{Blinc1978, Kitzerow2001} The amplitudon corresponds to a fluctuation in the magnitude of the tilt angle $\Delta \vartheta(t)=\vartheta-\vartheta_0$, whereas the phason represents a fluctuation in the direction of the tilt vector, \textit{i.e.}, a fluctuation in the phase $\phi=\phi(t)$ of the modulation wave. The SmA*-SmC* transition results from a soft-mode behaviour of the amplitudon. The phason corresponds to the Goldstone mode. It recovers the full rotational symmetry around the helical axes, which is broken at the para-to-ferroelectric phase transition due to collective tilts of the chiral mesogens. \cite{Kitzerow2001}

As discussed in the introduction the application of a large static electric field $\vec{E}$, in-plane of the smectic layer, forces an alignment of each layer's polarization vector $\vec{P}$ parallel to $\vec{E}$.\cite{Kitzerow2001} Consequently the molecular director of each individual smectic layer becomes perpendicular to the direction of the electric field $\vec{E}$ and the helical structure no longer exists.\cite{Kitzerow2001} Small static electric fields only slightly deform the helix structure, whereas alternating transverse fields excite eigenmodes of the helices, among which the phason mode are in the focus of electro-optical experiments. 

\begin{figure}[!b]
  \centering
  \includegraphics[angle=0,width=0.95\columnwidth]{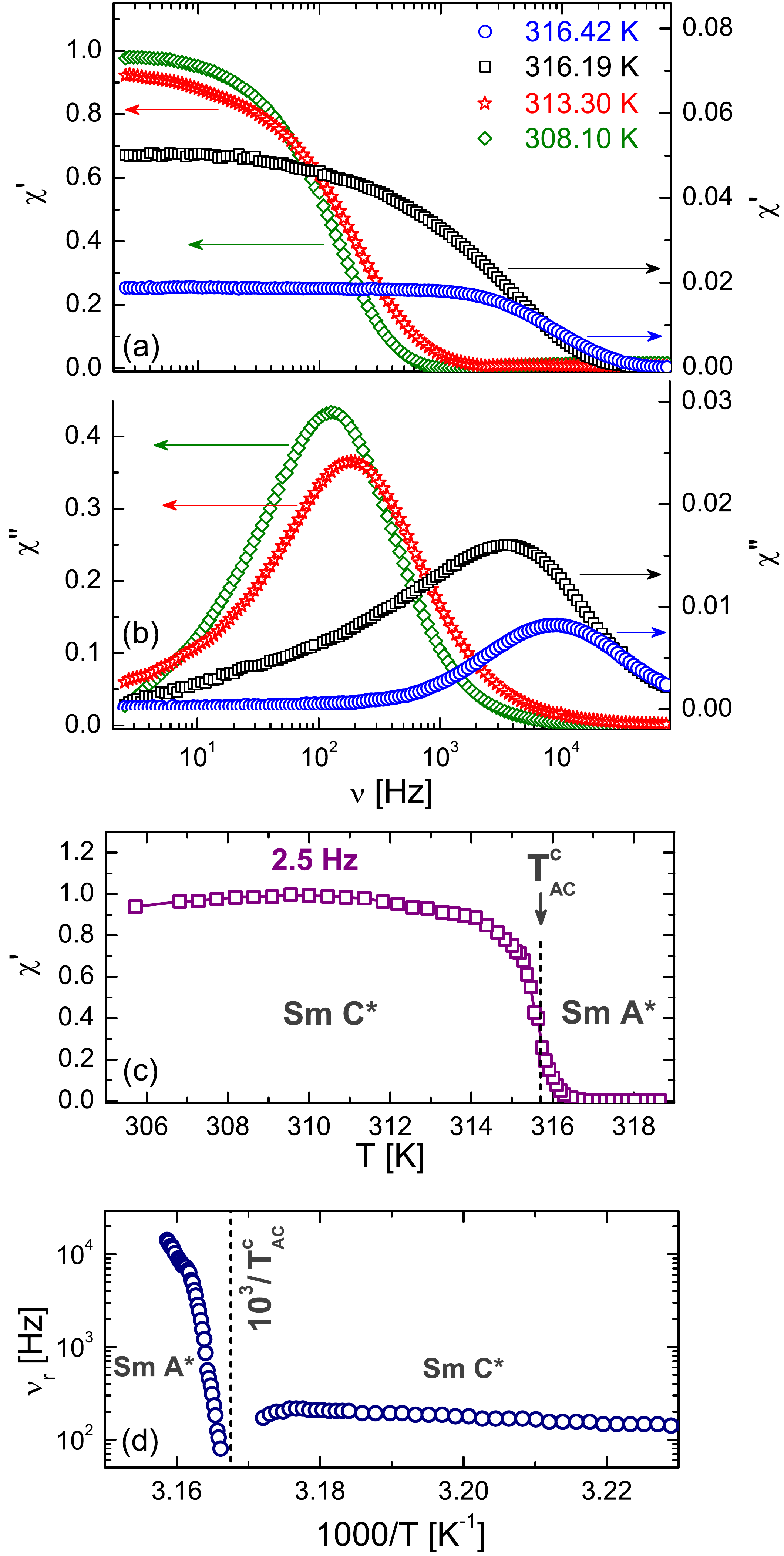}
  \caption{Real $\chi'$ (a) and imaginary $\chi''$ (b) parts of the normalised linear electro-optical response versus frequency, $\nu$, for the bulk FLC measured at several temperatures in the SmA* and SmC* phases. (c) Quasi-static linear electro-optical response $\chi'$ ($\nu=2.5$~Hz$ \ll \nu_r$) vs. $T$, being normalised to its maximum value. (d) Soft ($T >T_{AC}^c$) and phason ($T <T_{AC}^c$) modes relaxation frequencies vs $T^{-1}$.}
  \label{fig:electro-optic_bulk_data}
\end{figure}

In Fig.~\ref{fig:electro-optic_bulk_data}a and b we show the frequency dependence of the real and imaginary parts of the electro-optical response, $\chi'$ and $\chi''$, measured in the bulk sample.
The relaxation mode observed in the SmA* phase, \textit{i.e.}\ at $T>T_{AC}^c$ represents the soft mode, whereas well below this temperature the dielectric spectra is represented by the low frequency Goldstone phase (phason) mode, \cite{Gouda1991} as reported in a number of experiments, see \textit{e.g.}\ Refs. \cite{Rozanski2005,Rozanski2005a,Skarabot1998a}. In the close vicinity of $T_{AC}^c$ the splitting of the soft mode in the amplitudon and phason modes is expected, but an experimental observation of this splitting likewise of the amplitudon is challenging both in dielectric spectroscopy and electro-optical response experiments. \cite{Gouda1991} The relaxation frequency, $\nu_r$, of the observed modes, determined as the frequency position of the maximum in the $\chi''(\nu)$-dependencies exhibits characteristic temperature variations, see Fig.~\ref{fig:electro-optic_bulk_data}d. Particularly, a strong temperature-dependence of the soft mode above $T_{AC}^c$ contrasts with the rather weak changes of the low frequency Goldstone mode observed below $T_{AC}^c$. The real part of the normalised electro-optical response, $\chi'$, measured in the Hz-region ($\nu=2.5$~Hz) corresponds to the quasi-static regime. Its temperature-dependence (see Fig.~\ref{fig:electro-optic_bulk_data}c) is characterised by a sudden increase of $\chi'$ at $T_{AC}^c$ upon cooling. The bulk electro-optical response in the entire range of the SmA* phase, except in close vicinity of $T_{AC}^c$, is weak. Compared to the SmC* phase it is two orders of magnitudes smaller.

\begin{figure}
  \centering
  \includegraphics[angle=0,width=0.85\columnwidth]{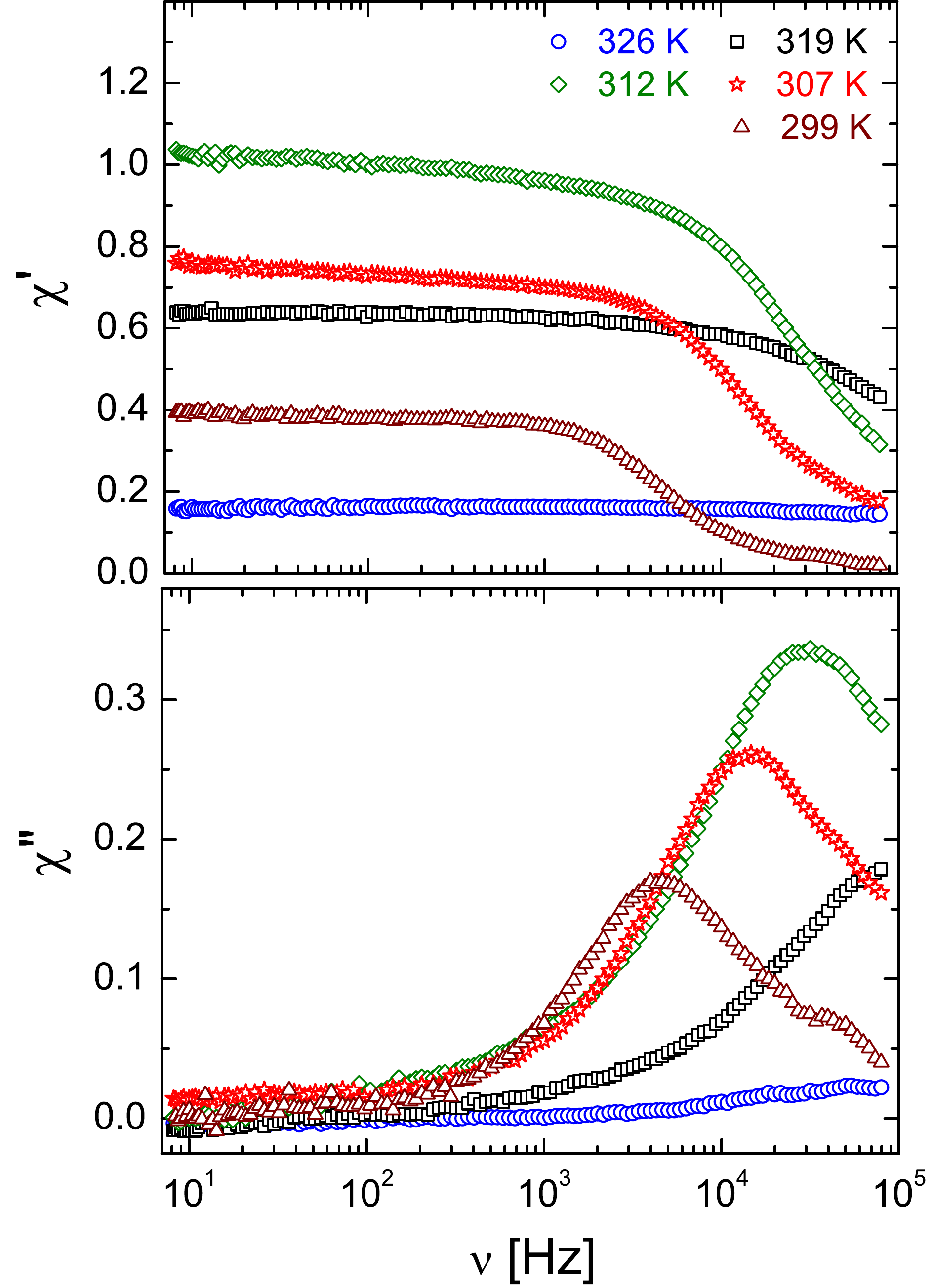}\hspace{-0.2cm}
  \caption{Real and imaginary parts of the linear electro-optical response, $\chi'$ and $\chi''$ vs frequency $\nu$ of the confined SmA* and SmC* phases for selected temperatures, confined in the $R=21$~nm membranes.}
  \label{fig:electro-optic_poresize_temp_freq_dependence}
\end{figure}

By contrast, the confined FLC shows an electro-optical effect also in the temperature range of the bulk SmA* phase, see Figs.~\ref{fig:Neutron_Intensity_layer-thickness_peak-width_electro-optic-response}d and \ref{fig:electro-optic_poresize_temp_freq_dependence}, but at low frequencies (in the quasi-static regime) it is several times smaller than the maximum value in the confined SmC* phase.
In the confined SmA* phase the characteristic relaxation frequencies, which would correspond to the soft relaxation mode, appear above our upper detection frequency. Depending on the pore diameter, the relaxation maxima in the $\chi''(\nu)$-dependence (see Fig.~\ref{fig:electro-optic_poresize_temp_freq_dependence} for $R=21.0$~nm and ESI$^\dag$ for other pore diameters) became observable below 308 to 314~K, \textit{i.e.}\ already well below the SmA*-to-SmC* transition, only.  Apparently it corresponds to the phason relaxation mode of the confined SmC* state. 

\begin{figure}[t]
  \centering
  \includegraphics[angle=0,width=0.95\columnwidth]{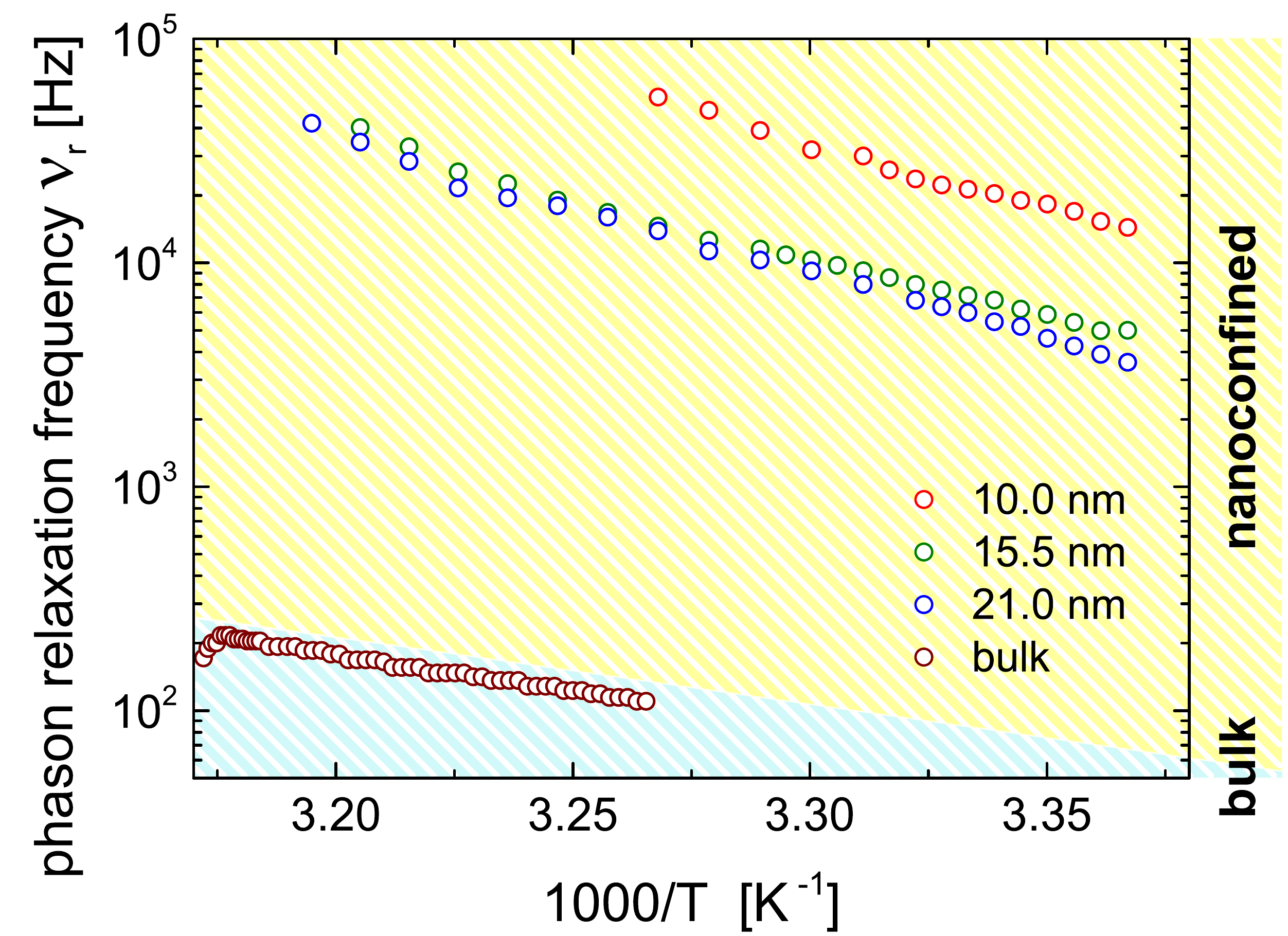}
  \caption{Temperature-dependence of the phason relaxation frequency, $\nu_r$ as Arrhenius plot for the bulk and the confined states as indicated in the figure.}
  \label{fig:phason_relaxation_frequency_arrhenius}
\end{figure}

The relaxation frequencies, $\nu_r$ vs $T^{-1}$, determined as position of the maxima in $\chi''(\nu)$-dependences are shown in comparison in Fig.~\ref{fig:phason_relaxation_frequency_arrhenius}.
Except for small deviations at higher temperatures they exhibit an Arrhenius-type behaviour $\nu_r = \nu^0_r \exp{(-E_a/(R_{\rm gas} T))}$ typical of a thermally activated process with activation energy $E_a$. The activation energies of the confined system, $E_a$=114$\pm$4~kJ/mol, 107$\pm$3~kJ/mol and 114$\pm$3~kJ/mol for the 10.0, 15.5 and 21.0~nm pores, respectively, are approximately twice the bulk value (63.0$\pm$0.8~kJ/mol), presumably because of the increased geometric and elastic constraints and thus higher energetic barriers for molecular rotations. In general, the temperature-dependence of the relaxation frequencies in the bulk and in the confined state can be traced to the temperature-dependence of the rotational viscosity originating from its thermally activated character.

An important conclusion about the character of the helical modulation in the confined SmC* phase can be achieved by analysing the relaxation frequency of the phason mode as a function of pore diameter. According to a Landau model analysis, the phason is a quasi-acoustic excitation with a resonance frequency $f_G \sim K_3 Q_G$, where $Q_G$ is the modulus of the excitation wave vector and $K_3$ the Frank bend elastic constant. \cite{Carlsson1990, Kitzerow2001} The corresponding relaxation time is $1/\tau_r \sim f_G^{-2}$, so that the relaxation frequency should scale with $\nu_r \sim K_3^2 Q_G^2$.  In dielectric or electro-optic measurements, as performed here, an electric field $\vec{E}$ is applied which excites a phason with a spatial characteristics determined by $\vec{E}$. In a bulk system $\vec{E}$ is homogeneous and the spatial period is given by the macroscopic extension $L$ ($L \rightarrow \infty$) of the sample. Thus, $Q_G \sim L^{-1}$ and the dielectric response corresponds to $Q_G$=0. In the thin-film- or pore confined state a non-homogeneous $\vec{E}$ field with a periodicity set by the wall distance $h$ or by the pore radius $R$ is applied, respectively, i.e. $Q_G \sim R$. Hence, $\tau_r$ should scale inversely proportional to the square of the length characterizing the spatial confinement.\\ 
In fact, previous theoretical and experimental studies, \cite{Skarabot1998a,Skarabot1999} performed with a wedge-type submicron cell (2D planar confinement), indicated the existence of two different regimes of structural modulation. Both a plane-wave and a soliton-like modulation can exist. In the two cases the phason relaxation frequency, $\nu_r$,  is characterised by a parabolic dispersion, as motivated above. However, the plane-wave dynamics in the limit of the inverse film height $h^{-1}\rightarrow 0$ is represented by a gapless excitation ($\nu_r \propto h^{-2}$), whereas the soliton-like modulation exhibits a finite gap, $g_o$, \textit{i.e.}\ corresponding to a dispersion relation $\nu_r \propto g_o + h^{-2}$. 

\begin{figure}[!b]
  \centering
  \includegraphics[angle=0,width=0.95\columnwidth]{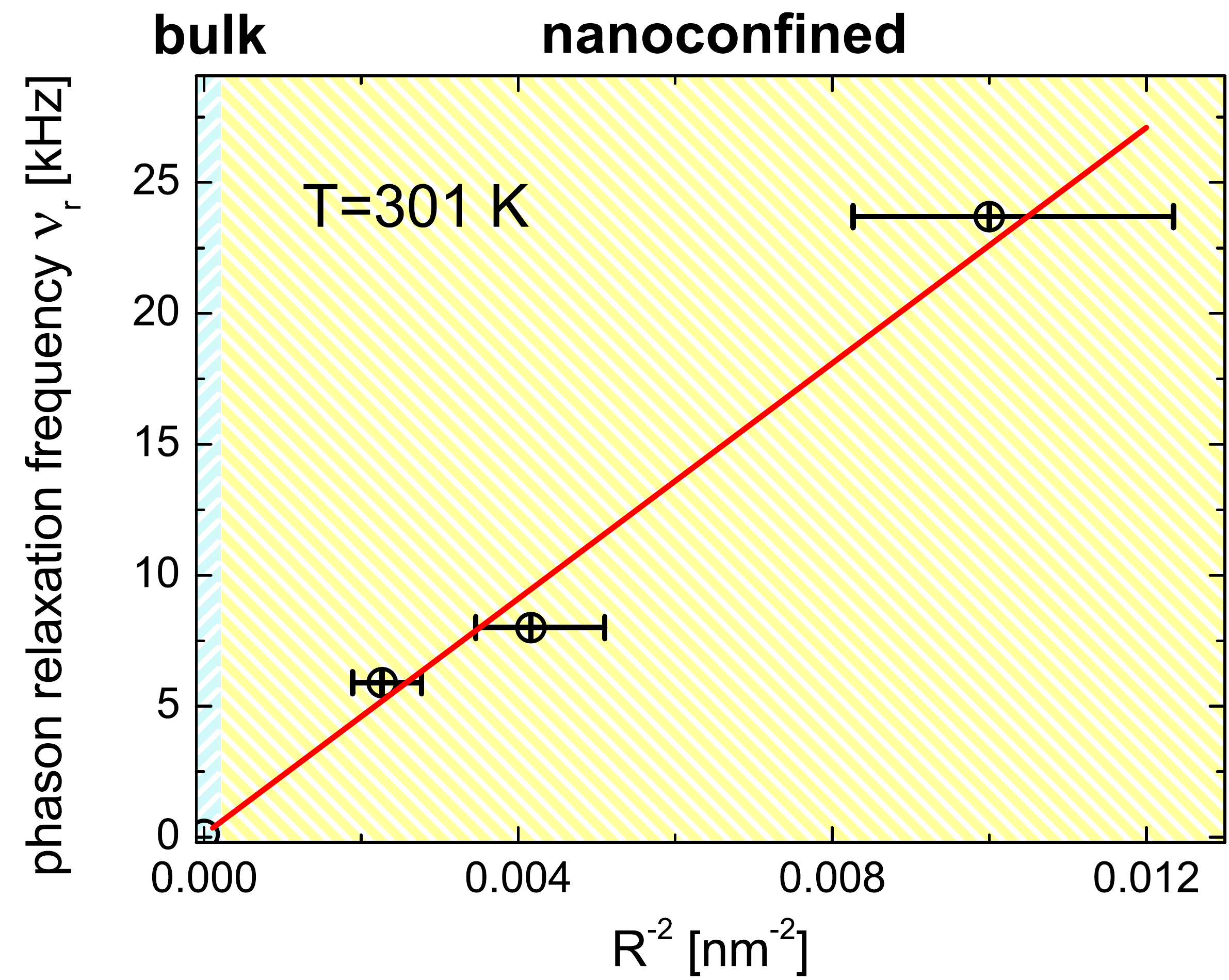}
  \caption{Phason relaxation frequency, $\nu_r$ vs. inverse nanopore radius squared for the confined FLC. Symbols are the experimental data points. The solid line is a linear fit.}
  \label{fig:phason_relaxation_frequency}
\end{figure}

In Fig.~\ref{fig:phason_relaxation_frequency} we plot the phase relaxation frequency $\nu_r$ versus the inverse pore radius squared, $R^{-2}$, at $T=301$~K. Within the error margins it is well fitted by a gapless $\nu_r(R^{-2})$-dispersion indicating that the nanoconfined helical structure in the SmC* phase exhibits a quasi-acoustic, plane-wave excitation characteristics. This finding is not too surprising, since the cylindrical geometry (2D-confinement) evidently suppresses the development of soliton-like distortions in any radial direction, \textit{i.e.}\ in any direction being perpendicular to the long pore axis. 

Note that an observation of a \emph{linear} electro-optical response in the confined SmC* phase corroborates the existence of the helical structure inferred above from the optical activity measurements. Moreover, nanoconfinement enhances the linear electro-optical response in the entire range of the confined SmA* phase, as can be seen in temperature-dependences of the quasi-static response, $\chi'(\nu_o=124$~Hz$)$ depicted in Fig.~\ref{fig:Neutron_Intensity_layer-thickness_peak-width_electro-optic-response}e for the confined FLCs. This observation corroborates our findings regarding the enhanced optical activity and the interpretation of the neutron diffraction experiments. Confinement induces a pretransitional, para-smectic state, which results in an enhanced linear electro-optical response both in the region of the SmC*$\rightleftharpoons$SmA* transition and, more importantly, in the entire temperature range of the bulk SmA* phase. 

$\chi'(T,\nu_o)$ (see Fig.~\ref{fig:Neutron_Intensity_layer-thickness_peak-width_electro-optic-response}e) exhibits another remarkable peculiarity. It shows a broad maximum in the confined SmC* phase, which shifts somewhat to lower temperatures with decreasing pore diameter. Hence, the behaviour of the normalised quasistatic electro-optical response in the nanoconfined geometry evidently differs from the one observed in the bulk. Two competing processes could explain such an electro-optical behaviour upon confinement. A considerable rise of the electro-optical response during cooling in the region of SmA*-to-SmC* is related to a formation of the helical structure due to the tilting of the molecules in smectic layers. In first approximation this effect is proportional to the spontaneous polarization accompanying the helix structure which in the region of this transition rises proportionally to the tilt angle and thus with decreasing temperature. By contrast, the orientational polarizability of the guest FLC molecules in an applied external electric field is hindered by interfacial interactions and by the rotational viscosity, which both increase with decreasing temperature. However, this effect is not strong enough to explain the steep decrease of the linear electro-optical response when cooling the confined FLC below approx.\ $T=309.5$~K (see Fig.~\ref{fig:Neutron_Intensity_layer-thickness_peak-width_electro-optic-response}e).  

In fact, this temperature agrees with the temperature where the neutron diffraction experiments indicate the gradual formation of chevron-like structures, see Fig.~\ref{fig:Neutron_Intensity_layer-thickness_peak-width_electro-optic-response}e. 
In contrast to the bookshelf structure, with a final electrical polarization perpendicular to the long pore axis within each layer, the rotational symmetry in the chevron-structures around the nanopore axis results in a significantly reduced or vanishing transverse electrical polarization. Thus the transverse external ac field $\vec{E}$ cannot excite any eigenmode of the helical structures and the electro-optical response should vanish in the extreme case of a full transformation towards chevrons. The reduced, but final electro-optical response at low temperature hints towards an incomplete transformation. Presumably there is a sizeable fraction of untilted layers in the pore center which is responsible for the final electro-optical response. This is also compatible with our neutron scattering experiments: Likely the two peaks in Fig.~\ref{fig:omega-scan-fits}a are accompanied by a third one at $\omega=0^\circ$. This would indicate untilted SmC* layers in the middle of the nanopores in coexistence with tilted layers at the pore circumference, see an idealised sketch in Fig.~\ref{fig:chevron-structure}b and explain the non-vanishing phason excitation at low temperature. 

\section{Conclusions}

We have presented temperature-dependent high-resolution optical birefringence, neutron diffraction and electro-optical experiments on soft-hard hybrid materials composed of mesoporous anodic aluminium oxide membranes with distinct pore radii and the ferroelectric liquid crystal \mbox{2MBOCBC}. The AAO membranes with native pore wall surfaces do not provide a stable molecular configuration for the confined guest molecules. By contrast, a thermal-history independent phase behaviour can be achieved by using polymer coatings, enhancing tangential molecular wall anchoring. It results in a formation of a smectic C* phase with high optical activity indicating the alignment of helical superstructures parallel to the long nanopore axes. 

The confined FLC exhibits a strongly increased light rotatory power compared to the bulk system. This enhancement can be traced to pretransitional effects at the smectic-A*-to-smectic-C* transformation, in particular the confinement-induced formation of supermolecular optically active states. The observation of a linear electro-optical effect allows us to study fluctuations in the molecular tilt vector direction along the helical superstructures, and thus phase fluctuations of the corresponding helicoidal modulation wave, \textit{i.e.}\ the so-called phason excitation. Its relaxation frequency increases with the squared inverse pore radii, is in the kHz regime and thus two orders of magnitude larger than in the bulk, evidencing a fast electro-optical response of the soft/hard nanocomposite. 

A sudden decay in this electro-optical response upon cooling below 310~K is traced to a partial transformation of smectic-C*-bookshelf to chevron-like structures, in excellent agreement with the neutron scattering experiment. This smectic layer buckling is attributed to the change in the layering distance at the SmC*-SmA* phase transition. Therefore, we envision studies of so-called ''de-Vries'' FLCs. They exhibit no or a neglegible layering-distance change upon entering into the SmC* phase. \cite{Lagerwall2006} Thus, for these systems the chevron-formation upon confinement should be suppressed.

We hope that our study will also stimulate theoretical studies. In particular simulations regarding the influences of restricted geometries on the supermolecular SmC* helices could result in a deeper understanding of the remarkable structural, dynamical and electro-optical properties found here, similarly as this has been achieved in the past for achiral and cholesteric LCs in nanoscale confinement. \cite{Gruhn1997, Care2005, Binder2008, Ji2009, Pizzirusso2012, Roscioni2013,Karjalainen2013, Cetinkaya2013, Schulz2014, Schlotthauer2015}

Finally, we believe that the robustness of the liquid-crystalline state in the nanoporous membranes and the evidences of exceptionally fast and strong, temperature-tunable electro-optic functionality are not only of fundamental importance for the understanding of the phase behaviour of chiral liquid-crystalline matter. Given the versatile tailorability of anodic aluminium oxide \cite{Chen2015} and other self-assembled, optically transparent mesoporous media \cite{Hoffmann2006, Corr2010, Kuster2014} it may also have technological benefits. For example, these soft-hard nanocomposites could be used as metamaterials, where the wavefront of light is controlled with subwavelength-spaced structures, opening up new opportunities to replace bulk optical devices, with thin, lightweight nanohybrids. \cite{Yao2014}

\section*{Conflict of interest}
There are no conflicts to declare.

\section*{Acknowledgements}

This work was supported by the Polish National Science Centre (NCN) 
under the project ''Molecular Structure
and Dynamics of Liquid Crystals Based Nanocomposites" (Decision no. 
DEC-2012/05/B/ST3/02782) and the Deutsche Forschungsgemeinschaft (DFG) within the Collaborative Research Initiative SFB 986, ''Tailor-Made Multi-Scale Materials Systems", project B7, project C2, Project Z3, Hamburg. We thank Dr.\ Thomas Hau\ss\ (Helmholtz-Zentrum Berlin) for supporting the neutron scattering experiments.



\balance


\bibliography{FLC_bibliography} 
\bibliographystyle{rsc} 

\end{document}